\documentclass[preprint]{elsart}
\usepackage{amssymb,amsmath}
\usepackage{units}

\usepackage{graphicx}
\usepackage{morefloats}

\usepackage{amsmath}
\usepackage{graphicx}
\usepackage{dcolumn}% Align table columns on decimal point
\usepackage{bm}% bold math
\usepackage[english,activeacute]{babel}

\begin{document}
%%%%%%%%%%%%%%%%%%%%%%%%%%%%%%%%%%%%%%%%%%%%%%%%%%%%%%%%%%%%%%%%%%%
\begin{frontmatter}

\title{A comprehensive study of 
shower to shower fluctuations}

\author[upl]{P. M. Hansen}
\author[usc]{J. Alvarez--Mu\~niz}
\author[usc]{R. A. V\'azquez}

\address[upl]{Departamento de Fisica, IFLP CONICET \\
Facultad de Ciencias Exactas, 
Universidad Nacional de La Plata C.C. 67,\\ 
1900 La Plata, Argentina}

\address[usc]{Departamento de F\'{i}sica de Part\'{i}culas \& \\
              Instituto Galego de F\'\i sica de Altas Energx\'\i as (IGFAE),\\
              Facultade de F\'{i}sica, Universidade de Santiago de Compostela,\\
              15782 Santiago de Compostela, Spain}

\begin{abstract}
By means of Monte Carlo simulations of extensive air showers (EAS), 
we have performed a comprehensive study of the shower to shower 
fluctuations affecting the longitudinal and lateral development of EAS.  
We split the fluctuations into physical fluctuations and 
those induced by the thinning procedure customarily 
applied to simulate showers at EeV energies and above. 
We study the influence of thinning on the calculation of the shower to shower
fluctuations in the simulations. For thinning levels larger than $R_{\rm thin}=10^{-5} - 10^{-6}$, 
the determination of the shower to shower fluctuations is hampered by the artificial 
fluctuations induced by the thinning procedure. However, we show that shower to shower fluctuations 
can still be approximately estimated, and we provide expressions to calculate 
them. The influence of fluctuations of the depth of first interaction on the
determination of shower to shower fluctuations is also addressed. 
\end{abstract}


\end{frontmatter}

\section{Introduction}

Extensive air showers (EAS) have been studied
over the last 70 years \cite{watson-nagano}.  They
result from the interaction in the atmosphere of high-energy protons 
and nuclei arriving from space.  The product of these collisions are a set of
secondary particles carrying a fraction of the primary energy.  These
secondaries move through the atmosphere and interact again 
generating new secondaries.  The process continues, increasing the number of
secondary particles, until their energies are too low to contribute to the 
generation of new particles. 
Particles reaching ground are sampled with
arrays of detectors, and their properties are used to infer the 
properties of the primary initiating the shower. Measurements of the
electron and muon density, of the arrival time of the particles at ground, 
and of the depth at which the shower has the maximum number of particles (X$_{\rm max}$), 
give information on the arrival direction, primary energy, and on the mass of the
primaries \cite{watson-nagano}. 

The complexity of the cascade phenomena, and the poor knowledge of the
hadronic interactions at very high energy \cite{hadronic_models}, make the experimental 
determination of the properties of the primaries very difficult. 
Moreover, primary particles with the same energy, mass and direction
produce secondary particles with parameters that vary from shower to
shower. This feature is called ``shower to shower fluctuations".  An
understanding of the shower to shower fluctuations will help to improve 
the interpretation of cosmic-ray data.

The calculation of shower to shower fluctuations can in principle
be addressed with Monte Carlo simulations of extensive air showers. 
However, the number of particles that are produced in an
air shower at ultra high energy (above $\sim 10^{18}$ eV) is so large ($\sim 10^{10}$),
that it is almost impossible to follow the propagation to ground level of all the secondaries in the Monte Carlo
in a reasonable amount of time, or even to store the large amount of information produced. 
For this reason, a statistical sampling procedure called ``thinning" \cite{Hillas_thin} is used in the simulations. 
Thinning algorithms typically consist on propagating only a small, representative
fraction of the total number of particles in the shower, assigning statistical weights 
to the sampled particles to compensate for the rejected ones.  
However, thinning algorithms introduce 
artificial fluctuations in the simulated showers, hampering the determination 
of the intrinsic, physical shower to shower fluctuations with Monte Carlo simulations.   
For this reason the study of fluctuations using Monte Carlo simulations 
is quite difficult and uncertain. 
This is of utmost importance in cosmic-ray physics, since an incorrect assumption 
on the shower to shower fluctuations can lead 
to systematic errors on the determination of the parameters of the primary
particles.  

In this work we address the problem of determining the true, physical shower to shower fluctuations in 
Monte Carlo simulations, and quantify the effect of thinning on their determination. 
We give expressions that allow the estimation of 
physical fluctuations from Monte Carlo simulations, 
even in the case of relatively strongly thinned showers. 

The paper is organized as follows: In Section \ref{S:TheSimulations} we
describe the simulations performed in this work, and the thinning algorithm adopted. In Section
\ref{S:Fluctuations} we identify the different sources of fluctuations
in shower simulations.  
In Section \ref{S:Results} we perform a comprehensive study of the 
fluctuations in the longitudinal and lateral shower development, and give
expressions that allow to separate physical shower to shower fluctuations
from the artificial fluctuations induced by the thinning procedure. 
In Section \ref{S:Composition},
we quantify the influence on the shower to shower fluctuations of the 
fluctuations of the depth of first interaction of the primary initiating the shower. 
Finally, we summarize our conclusions in section \ref{S:Conclusions}. In the
Appendix we give an explicit mathematical derivation of the expressions presented in 
Section~\ref{S:Results}. 

\section{The simulations}
\label{S:TheSimulations}
In this work we have used the air shower simulation program, AIRES
\cite{AIRES,AIRESManual}, along with the hadronic model QGSJET01 \cite{QGSJET} to
simulate proton and iron-induced showers with primary energy $10^{19}$ eV.
As explained above, due to the large number of particles that are created in the simulation, 
AIRES includes a statistical sampling algorithm, that
consists on propagating a small, representative fraction of the total number of
particles, assigning a statistical weight $w$ to the sampled
particles to compensate for the rejected ones.  The weight is adjusted in such
a way that both the total energy and the average number of particles is
guaranteed to be conserved.

Before the simulation starts, the user indicates, as an input to AIRES, the relative
thinning level $R_{\rm thin}$. The thinning energy $E_{\rm thin}$ - the energy
below which the thinning process starts - is defined as $E_{\rm thin}$=$R_{\rm thin} \times
E_{p}$ where, $E_{p}$ is the primary energy. For ultra high energy cosmic ray
shower simulations convenient values for the relative thinning are $R_{\rm thin}
=10^{-5} - 10^{-9}$, but the actual choice depends on the purpose of the simulation.
The thinning level affects both the simulation 
CPU time and the size of the output produced in the simulation, both typically 
behaving linearly with $R_{\rm thin}^{-1}$. If 
we increase $R_{\rm thin}$ by a factor of 10 the simulation speeds up by a similar
factor, and the output is reduced accordingly, but the
price to pay is an enhancement of the artifical fluctuations in the simulated 
showers as discussed below.

We describe here the thinning algorithm implemented in the AIRES 
code \cite{AIRESManual}, originally due to Hillas \cite{Hillas_thin}.  At the beginning of the
simulation, the primary particle is assigned a weight $w=1$. Then the primary
is propagated and interacts in the atmosphere producing $n$ secondary particles.
Before incorporating any secondary particle in the simulation,
the energy of the primary $E_p$ which has generated that secondary is
compared to $E_{\rm thin}$.  If $E_p >  E_{\rm thin}$, then all the
secondaries with energy greater or equal than $E_{\rm thin}$ are kept, and their
weight is equal to the weight of the primary particle. Secondaries with energy less than
$E_{\rm thin}$ are kept with a probability $p_i = E_{i}/E_{\rm thin}$ ($E_i$ is the energy
of $i^{\rm th}$ secondary), and their weight is adjusted so that $w_i = (1/p_i) \times w$, with 
$w$ being the weight of the mother particle producing that secondary. 
On the other hand if $E_p < E_{\rm thin}$, it means that
the particle came from a previous thinning operation. Then, one and only one of all
the produced secondaries - say the $j^{\rm th}$ -  is kept, with probability
$p_{j}=E_j/\sum \limits_{i=1}^{n} E_{i}$. 
Again, the weight of this particle is
increased by a factor $w_{j}=(1/p_{j})\times w$.

To avoid confusion between particles and weights, we identify an {\it entry}
with a particle explicitely followed in the simulation which is associated a weight $w$. 
Hence, an entry represents $w$ particles.
It is important to stress that once the thinning energy is reached, the number
of entries $N_e$ is no longer increased in the shower processes (only one
secondary particle is followed in each interaction), while the
number of particles $N$ does however increase, since the weight of each entry 
typically increases in the showering process.
When evaluating a physical observable, 
each entry must be weighted with its corresponding statistical weight.

In AIRES, the thinning algorithm is complemented with an ``extended thinning
algorithm'' \cite{AIRESManual}, designed to ensure that all the statistical
weights are always smaller than a certain positive number (other algorithms
based on this same idea are possible, see for instance \cite{Kobal}). 
To ensure this, an external parameter called statistical weight factor $W$ is available
in the simulation. 
To further optimize the procedure of sampling,
separated weight factors for electromagnetic ($W({\rm EM})$) and heavy
particles ($W({\rm HADRONIC})$), are defined.  The parameter $W({\rm
  HADRONIC})$ is specified indirectly by the ratio:
\begin{equation}
{\rm AEH}=\frac{W({\rm EM})}{W({\rm HADRONIC})}.
\label{equ:fit}
\end{equation}
The default value of this ratio in the simulation is ${\rm AEH}=88$.  The
default value of the weight factor $W({\rm EM})$ is 12. In this paper, we will
use these default values unless otherwise specified.

We have simulated proton and iron-induced showers with primary energy $E_p=10^{19}$ eV,
zenith angle $\theta=0^\circ,~30^\circ,~45^\circ$ and $60^\circ$ 
and relative thinning $R_{\rm thin}=10^{-5},~10^{-6},~10^{-7}$ and $10^{-8}$
(the two latter $R_{\rm thin}$ for proton only).
We have also simulated showers with relative
thinning of $R_{\rm thin}=10^{-7}$ but with $W({\rm EM})= 0.1$
instead of the default value.
Finally, we have also simulated two sets of proton and iron-induced 
showers at $\theta=0^\circ$, with fixed depth of first
interaction, starting at the corresponding mean interaction depth
for protons and iron at $10^{19}$ eV.

\section{Fluctuations in EAS}
\label{S:Fluctuations}
In a real shower or in a simulation of an EAS, there are a number of different 
fluctuations that can occur. Rather generally, we can make a simple classification as 
shown in Table \ref{table:exp}.

``Physical fluctuations'' are those due to physical processes in the shower. 
Here we split them into those due to the first interaction, and those
occuring in the secondary interactions, as is customary, and because 
it has recently been suggested that ``universal'' shower properties may emerge when 
considering only the fluctuations in the first interaction point
\cite{universality}. Physical fluctuations occuring in the first interaction 
are further divided into those affecting the depth of the first interaction, 
and those that arise from fluctuations of multiplicity or inelasticity
also in the first interaction. 

\begin{table}[htbp]
\begin{tabular}{ l l }
\hline
Physical fluctuations  & ~~~~~ - Depth of first interaction.  \\
                       &  ~~~~~ - Multiplicity, inelasticity, etc, in $1^{\rm st}$ interaction.  \\
                       &  ~~~~~ - Secondary interactions.  \\
\hline
 Experimental fluctuations & ~~~~~ - Detector response.  \\
                           & ~~~~~ - Sampling fluctuations.  \\
\hline
 Artificial fluctuations & ~~~~~  - Thinning.  \\
                         & ~~~~~  - Un-thinning.  \\
\hline
\end{tabular}
\vspace{0.5cm}
\caption{Classification of the fluctuations in a shower, arising 
from the physical processes in the shower and the measurement process, and
those that appear only in shower simulations.}
\label{table:exp}
\end{table}

In the case of real data, fluctuations are enlarged due to the detector
response, and to the fact that the detector usually only samples a small
fraction of the shower front. This ``sampling fluctuation" is a statistically well
known problem, and sampling fluctuations are rather well studied \cite{Billoir,bauleo}. We will not consider
them in this work. Also the detector response
introduces an additional source of fluctuations, which are detector dependent,
and will not be considered here.

On the other hand, Monte Carlo simulated data is affected by artificial fluctuations due
to the thinning and un-thinning (re-sampling) procedures. 
%These affect the interpretation of
%Monte Carlo simulations and make them difficult to be usable in the generation
%of artificial events, for instance. 
For the purposes of this work we do not
need to consider the effect of fluctuations induced by the unthinning procedure \cite{Billoir}. 
%since there is not a standard procedure  
%and the theory behind it is not yet very clear.

\section{Fluctuations of the longitudinal and lateral shower development}
\label{S:Results}

\subsection{Fluctuations of the longitudinal profile}

\begin{figure}[p]
{\centering \begin{tabular}{cc}
\resizebox*{0.49\textwidth}{!}{\includegraphics{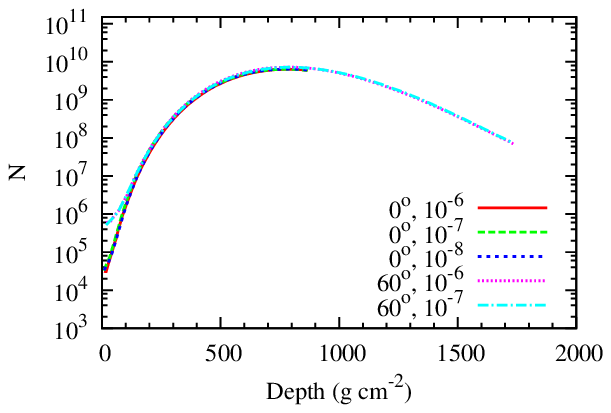}}
&\resizebox*{0.49\textwidth}{!}{\includegraphics{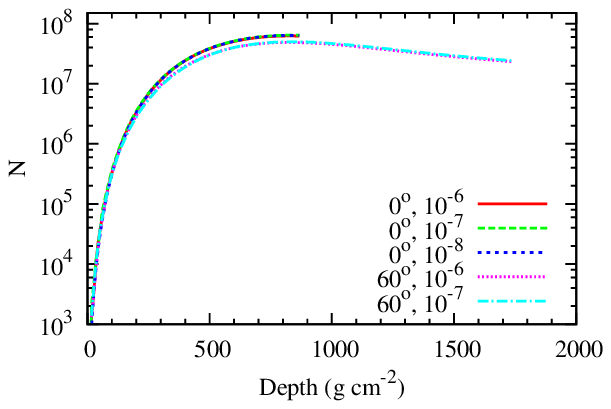}}
\\
\resizebox*{0.49\textwidth}{!}{\includegraphics{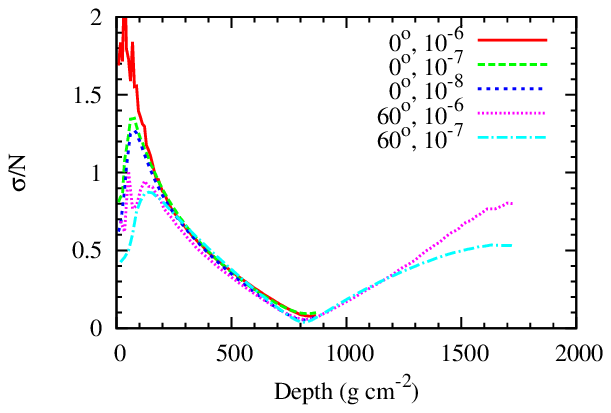}}
&\resizebox*{0.49\textwidth}{!}{\includegraphics{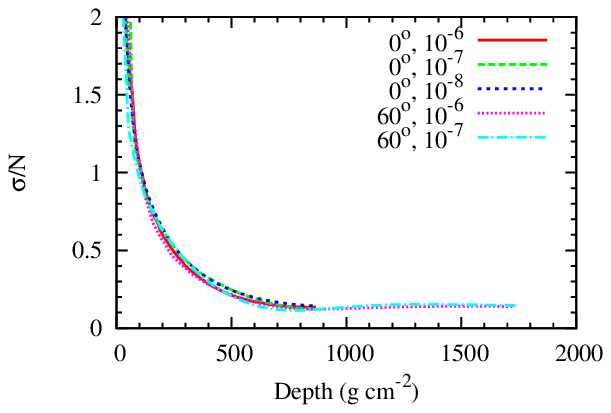}}
\\
\end{tabular}\par}
\caption{Upper panels: Longitudinal development of the average number of electrons (left)
and muons (right) as a function of
the slanted depth. 
Lower panels: Relative fluctuations $\sigma/\bar N$ as a function of
the slant depth for electrons (left) and muons (right). In all panels:
100 proton-induced showers with $E_p=10^{19}$ eV and $\theta=0^\circ$ and $60^\circ$, 
were simulated with relative thinning $R_{\rm thin}=10^{-6}$, 10$^{-7}$ and 10$^{-8}$. }
\label{fig:longitudinal}
\end{figure}

In Fig.~\ref{fig:longitudinal} we show the average longitudinal profile of
the number of electrons (left panels) and muons (right panels) $\bar N$, obtained in simulations of 
100 proton-induced showers with $E_p=10^{19}$ eV, for thinning levels $R_{\rm thin} = 10^{-6}$ and
$10^{-7}$ and $\theta=0^\circ,~60^\circ$. Also shown are the relative
shower to shower fluctuations $\sigma/\bar N$ for electrons and muons. As it is well known \cite{Gaisser},
the relative fluctuation has a minimum close to the depth of shower maximum.  
Also and as it is apparent from the figures, the dependence of the relative 
fluctuation on the thinning is small, at
least for depths close to the depth of maximum.

\begin{figure}[p]
{\centering \begin{tabular}{cc}
\resizebox*{0.49\textwidth}{!}{\includegraphics{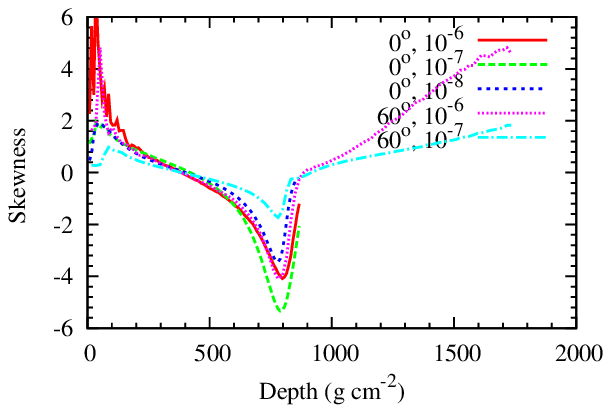}}
&\resizebox*{0.49\textwidth}{!}{\includegraphics{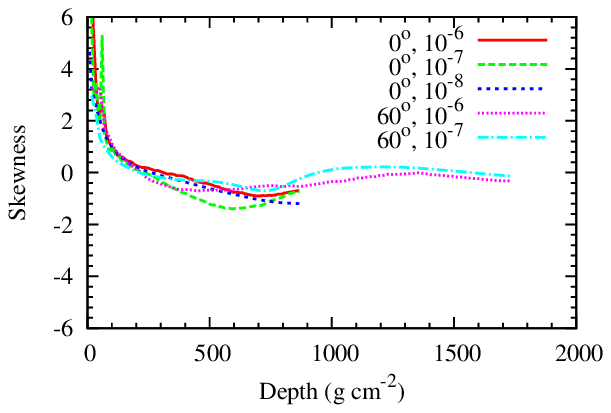}}
\\
\resizebox*{0.49\textwidth}{!}{\includegraphics{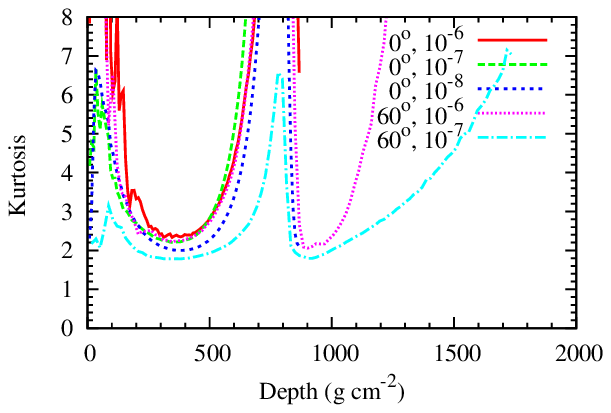}}
&\resizebox*{0.49\textwidth}{!}{\includegraphics{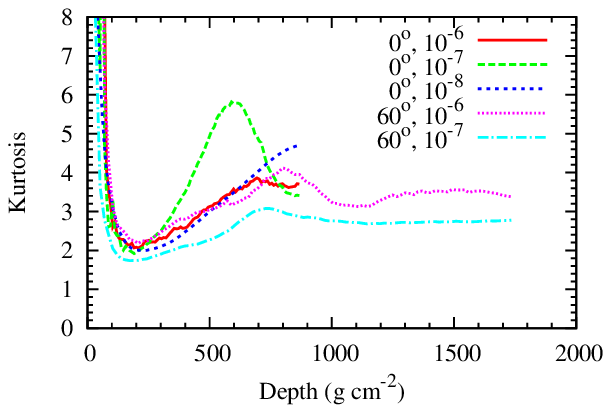}}
\\
\end{tabular}\par}
\caption{Upper panels: Skewness of the distribution of number of 
electrons (left) and muons (right) as a function of
the slant depth.  
Lower panels: Same as the upper panels for the kurtosis.
In all panels:
100 proton-induced showers of $E_p=10^{19}$ eV with $\theta=0^\circ$ and $60^\circ$ 
were simulated with 
relative thinning $R_{\rm th}=10^{-6}$, 10$^{-7}$ and $10^{-8}$.}
\label{fig:moments_long}
\end{figure}

In Fig.~\ref{fig:moments_long} we show for the same showers in Fig.~\ref{fig:longitudinal}, 
the skewness and the kurtosis of the distribution of the number of particles $N$
at different depths. It is worth recalling that
the skewness of the distribution of a variable $x$ is defined as
\begin{equation}
\gamma_3 = \frac{\langle (x-\bar x)^3\rangle}{\sigma_x^3},
\end{equation}
where $\bar x$ ($\sigma_x$) is the average (standard deviation) of $x$. 
The kurtosis is defined as 
\begin{equation}
\gamma_4 = \frac{\langle (x-\bar x)^4\rangle}{\sigma_x^4} -3.
\end{equation}
where the ``$-3$" in the definition is a convention to make $\gamma_4 = 0$ for a
Gaussian distribution. Both the kurtosis and the skewness can be positive or
negative. The skewness is a measure of the asymmetry of the distribution with respect
to the mean value. A negative sign implies that the distribution is ``deformed"
towards values of $x$ smaller than the mean. The contrary applies for a positive sign. 
The kurtosis is a measure of the length of the tails of the distribution.  Positive values
imply that the distribution has tails longer than those of a Gaussian, while negative
values imply that the tails are shorter  (for instance a flat
distribution, a box, has kurtosis -1.2). A Gaussian distribution has
$\gamma_3 = \gamma_4 = 0$. 

Several remarks can be made from Fig.~\ref{fig:moments_long}.
Firstly, it is apparent that both, the skewness and the kurtosis of the
distribution of the number of particles depend strongly on the
thinning level, contrary to what happens to the mean $\bar N$ and to the relative 
fluctuations $\sigma/\bar N$. 
Close to the depth of shower maximum both the skewness and the kurtosis have local
extrema very different from zero, implying that the distribution is strongly non-Gaussian. 
The skewness is negative and this implies that the distribution is
asymmetric towards smaller values of $N$ than average. The positive values 
of the kurtosis imply that the distribution of $N$ has tails longer than those of 
a Gaussian, at least close to shower maximum. Remarkably, the log-Gaussian 
distribution, widely used to parameterize fluctuations in the number of electrons, 
has both $\gamma_3>0$ and $\gamma_4>0$, while the fluctuations predicted by Monte Carlo
simulations near the maximum of the shower, have negative skewness.
For muons and at large depths
the skewness is close to zero, so that a Gaussian or a 
log-Gaussian distribution is a good approximation. For electrons, this is never the case.

\subsection{Fluctuations of the lateral profile at ground}

Of special importance for cosmic-ray physics  
performed with arrays of detectors is 
the study of fluctuations in the number of particles at ground.

\begin{figure}[p]
{\centering \begin{tabular}{cc}
\resizebox*{0.49\textwidth}{!}{\includegraphics{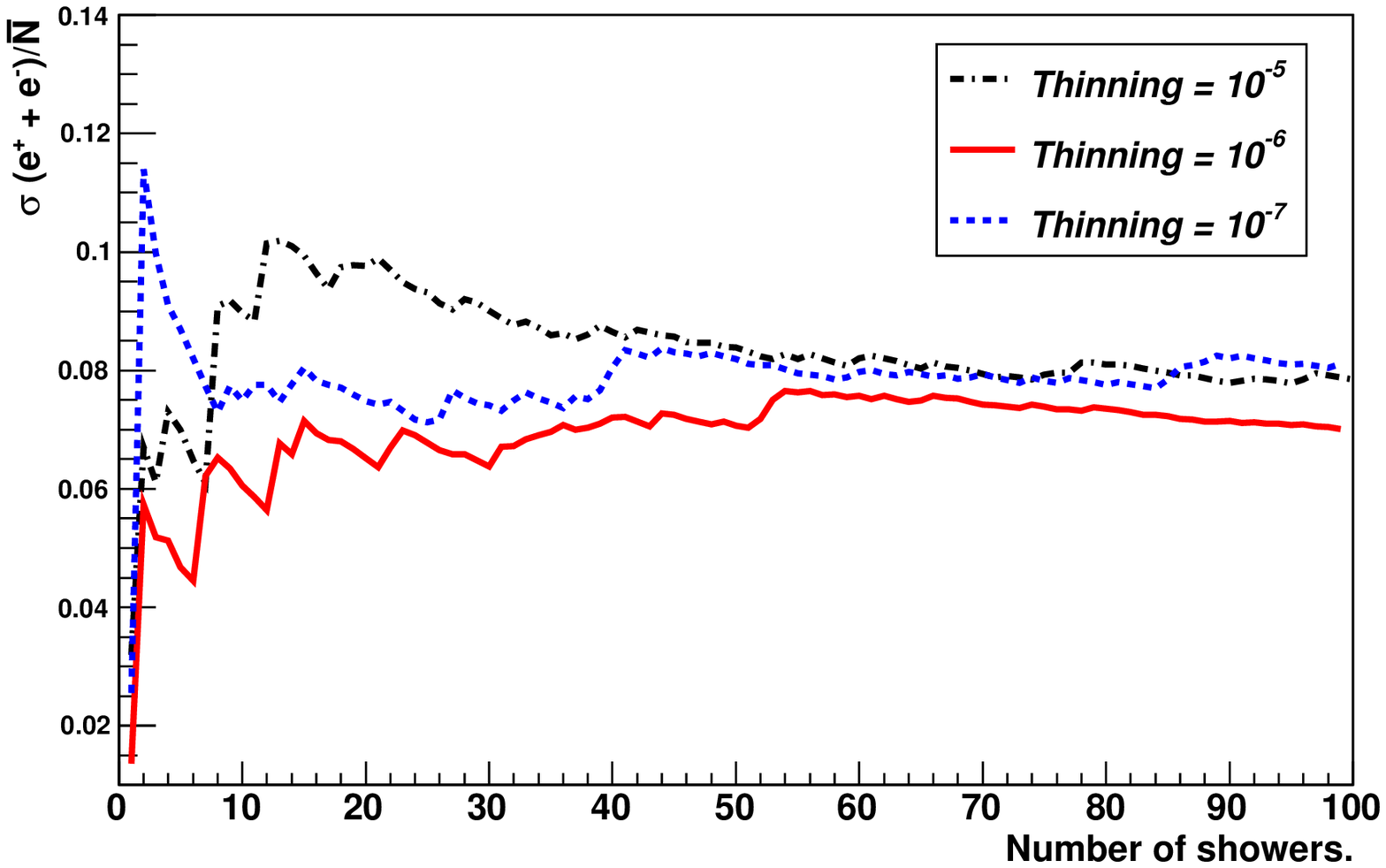}}
&\resizebox*{0.49\textwidth}{!}{\includegraphics{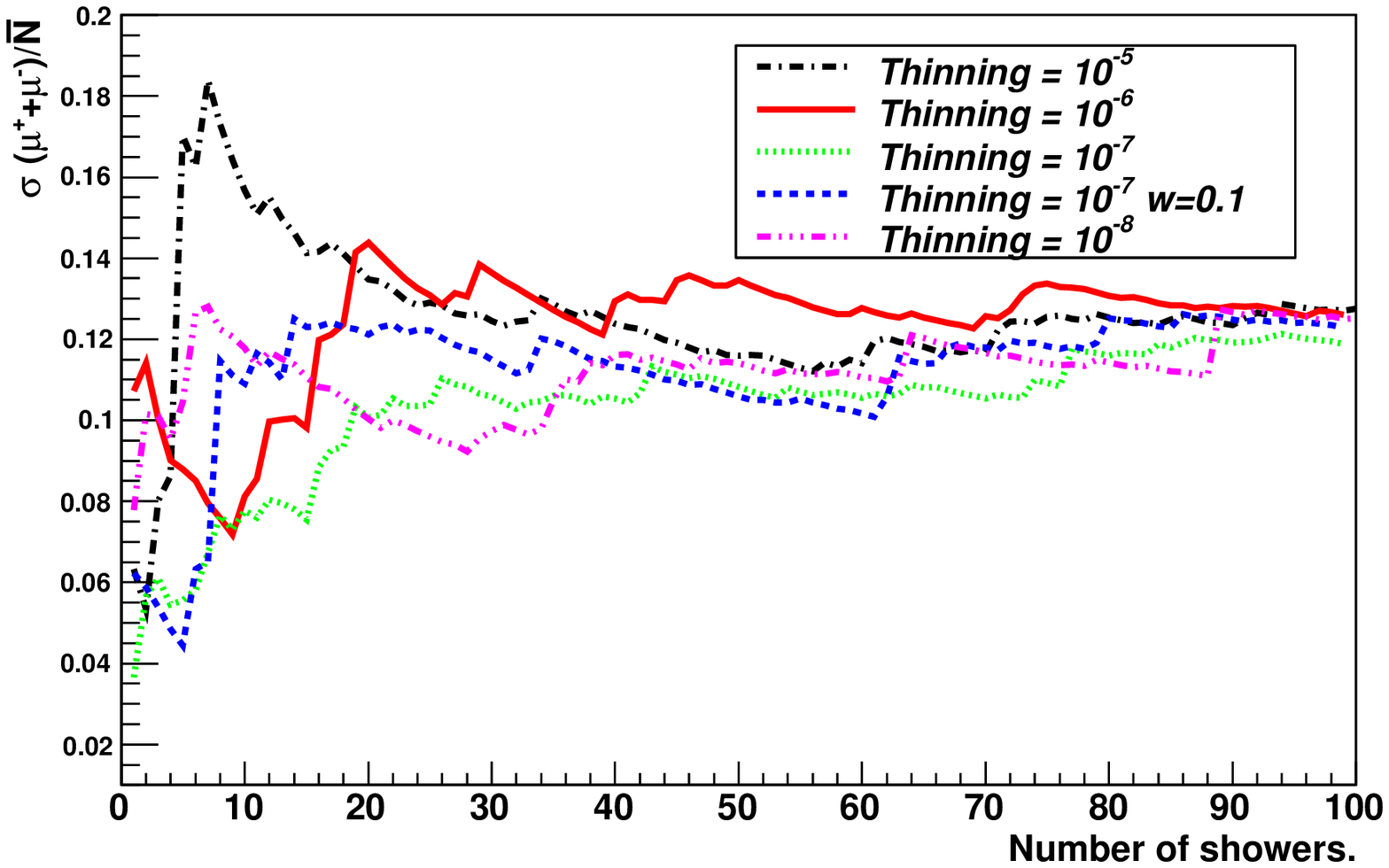}}
\\
\resizebox*{0.49\textwidth}{!}{\includegraphics{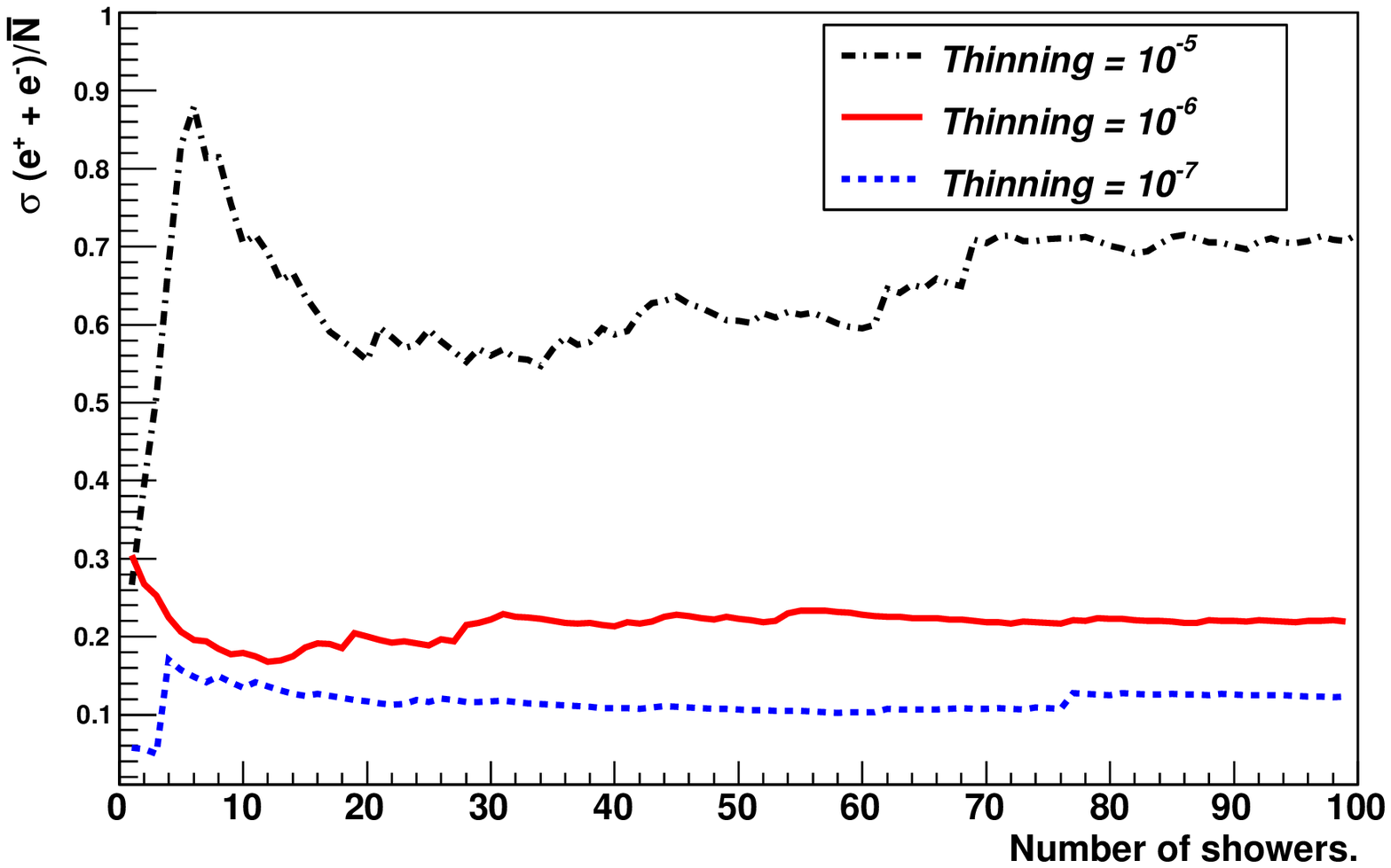}}
&\resizebox*{0.49\textwidth}{!}{\includegraphics{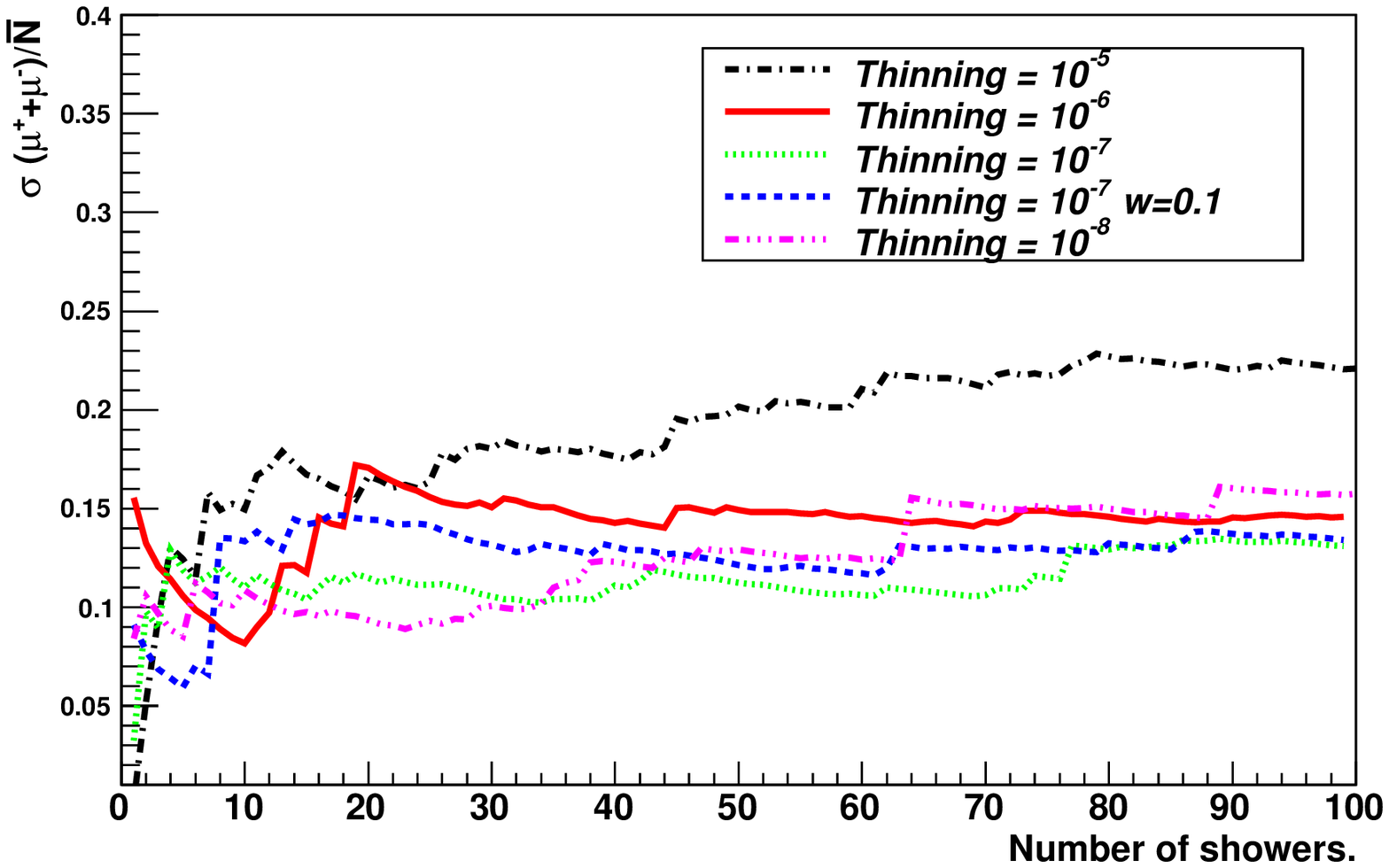}}
\\
\end{tabular}\par}
\caption{Relative fluctuations ($\sigma/\bar N$) in the number of electrons
(left panels) and muons (right panels) 
  versus the number of showers in the simulations.  
  In the upper panels we show the relative fluctuation of the total number of
  particles at ground.  In the lower panels we show the
  fluctuations of the number of particles falling in a ring $\Delta r$ at ground around
  a distance to the shower core $r=1000$ m. In all panels: proton-induced showers 
  of energy $E_p=10^{19}$ eV and $\theta=0^\circ$ were simulated with relative
thinning as indicated in the insets.}
\label{fig:fig1}
\end{figure}

In Fig.~\ref{fig:fig1} we show the relative fluctuations ($\sigma/\bar N$) 
of the total number of electrons (left panels) and muons (right panels) at ground 
(upper panels), and in a ring of width $\Delta r$ at a distance $r=1000$ m 
from the shower axis (lower panels). In both cases the fluctuations are
shown as a function of the number of showers simulated. The ring was taken from 
$r_{\rm min}=912$ m to $r_{\rm max}=1092$ m, i.e. $\Delta r=180$ m corresponding
to a symmetric interval in the logarithm of $r$ around $r=1000$ m, chosen so 
that it compensates the decreasing density of particles 
with a larger area as $r$ increases. 
As expected, fluctuations in the ring $\Delta r$
are larger than the fluctuations in the whole ground. 
Also, the fluctuations in the ring have a stronger dependence with the
thinning level used than those in the whole ground. This is easy to understand.
An entry of weight $w$ falling in the ring represents 
$w$ particles, so that by losing or
gaining just a single entry, one would lose or gain $w$ particles and the 
fluctuations are enlarged. This effect is not so strong when accounting for
all the particles falling anywhere on the ground.  
In Fig.~\ref{fig:fig1} it can also be clearly seen that no reliable evaluation of the fluctuations
can be done with less than about 20 simulated showers, especially in the case of 
fluctuations in the ring. It can also be seen that a thinning level of $R_{\rm thin}=10^{-6}$ 
or larger, introduces large artificial fluctuations, so that   
the shower to shower physical fluctuations cannot be evaluated reliably.  
This is however no obstacle to approximately estimate the physical shower to shower
fluctuations as will be shown in the following. 

\subsubsection{Physical shower to shower fluctuations at ground}

In the Appendix we prove that the distribution of the number of particles $N$ 
as obtained in thinned Monte Carlo simulations of extensive air showers has a mean 
$\bar N$ and a standard deviation $\sigma$ given by:
\begin{equation}
\bar N   =  \bar N_e \; \bar w,  
\label{eq:mean}
\end{equation}
and,
\begin{equation}
\sigma^2 =  \bar N_e \Omega^2 + \bar w^2 s^2. 
\label{eq:sigma}
\end{equation}
where:
\begin{itemize}

\item 
$\bar N_e$ and $s$ are respectively the mean and the standard deviation of
the distribution of the number of entries $N_e$
falling in a given ring around shower axis in each shower, 
i.e., the distribution 
of the number of non-thinned (explicitely sampled) particles in the simulation.

\item
$\bar w$ and $\Omega$ are respectively the mean and the standard deviation
of the distribution of weights assigned to the entries.

\end{itemize}

Of course Eq.~(\ref{eq:mean}) is exact since the thinning algorithm
is designed to reproduce it. For Eq.~(\ref{eq:sigma}), the proof
only assumes that the probability for an entry to have a given weight $w$ is
independent of the probability of a shower to have a given number of entries $N_e$.
This is only approximate since the total number of entries and their
weights are constrained by energy conservation.

One can interpret Eq.~(\ref{eq:sigma}) as follows. If all entries (sampled particles) 
had the same weight equal to $\bar w$ in all the simulated showers, then the distribution of 
weights would not fluctuate
from shower to shower and we would get $\Omega = 0$. In this limit we would obviously have $\sigma = \bar w s$. 
This special case of Eq.~(\ref{eq:sigma}) was also found in \cite{Knapp}. 
In the particular case in which all weights are equal to $\bar w=1$, i.e. the shower is fully simulated
and the thinning procedure is not applied, then clearly $N=N_e$, $\sigma = s$ and the fluctuations would be
obviously dominated by the true, physical shower to shower fluctuations.  
In the opposite limit, if we imagine that showers always have the same
number of entries equal to $\bar N_e$, i.e. there are no physical shower to shower 
fluctuations, then we would get $s=0$, and the fluctuation 
in the number of particles
would be solely due to the fluctuations of the weight of the entries, i.e., the fluctuations
would be dominated by the thinning procedure and  
$\sigma^2 = \bar N_e \Omega^2$. Therefore, we can identify the first term of
Eq.~(\ref{eq:sigma}) with the artificial fluctuations introduced by the thinning procedure, 
and the second term with the true (physical) shower to shower fluctuations and define: 
\begin{equation}
\sigma^2_{\rm thin} = \bar N_e \Omega^2,
\end{equation}
and
\begin{equation}
\sigma^2_{\rm phys} = \bar w^2 s^2.
\end{equation}

Eq.~(\ref{eq:sigma}) allows us to split the fluctuations into artificial and true
ones and hence to estimate the effect of the thinning in a particular set of simulations, performed
even with a relatively large value of the thinning level $R_{\rm thin}\sim 10^{-6}$, 
without the need to run new, more time-consuming and sometimes even impractical 
simulations with smaller $R_{\rm thin}$.

In the following, and by means of our Monte Carlo simulations, we numerically 
verify two key elements. Firstly, that Eq.~(\ref{eq:sigma}) accounts for all the
fluctuations (artificial and physical) appearing in simulations of EAS
with thinning; and secondly that as $R_{\rm thin}\rightarrow 0$, (and  
the effect of the thinning procedure on the fluctuations is decreasingly 
important so that the fluctuations are increasingly dominated by the physical shower to 
shower fluctuations), the second term in Eq.~(\ref{eq:sigma}) suffices to describe 
the fluctuations obtained in the Monte Carlo simulations. 

To see this in detail, we have calculated in Monte Carlo simulations, 
the average weight $\bar w$ and 
the sigma of the distribution of weights $\Omega$, the average number of entries
$N_e$ and the corresponding sigma of its distribution $s$, as well as 
the average number of particles $\bar N$ and the sigma of its distribution 
$\sigma$, both  for electrons and muons, and compared them to what is 
predicted by Eq.~(\ref{eq:sigma}). 
  
\begin{figure}[ht]
{\centering \begin{tabular}{cc}
\includegraphics[angle=0, width=7.0cm]{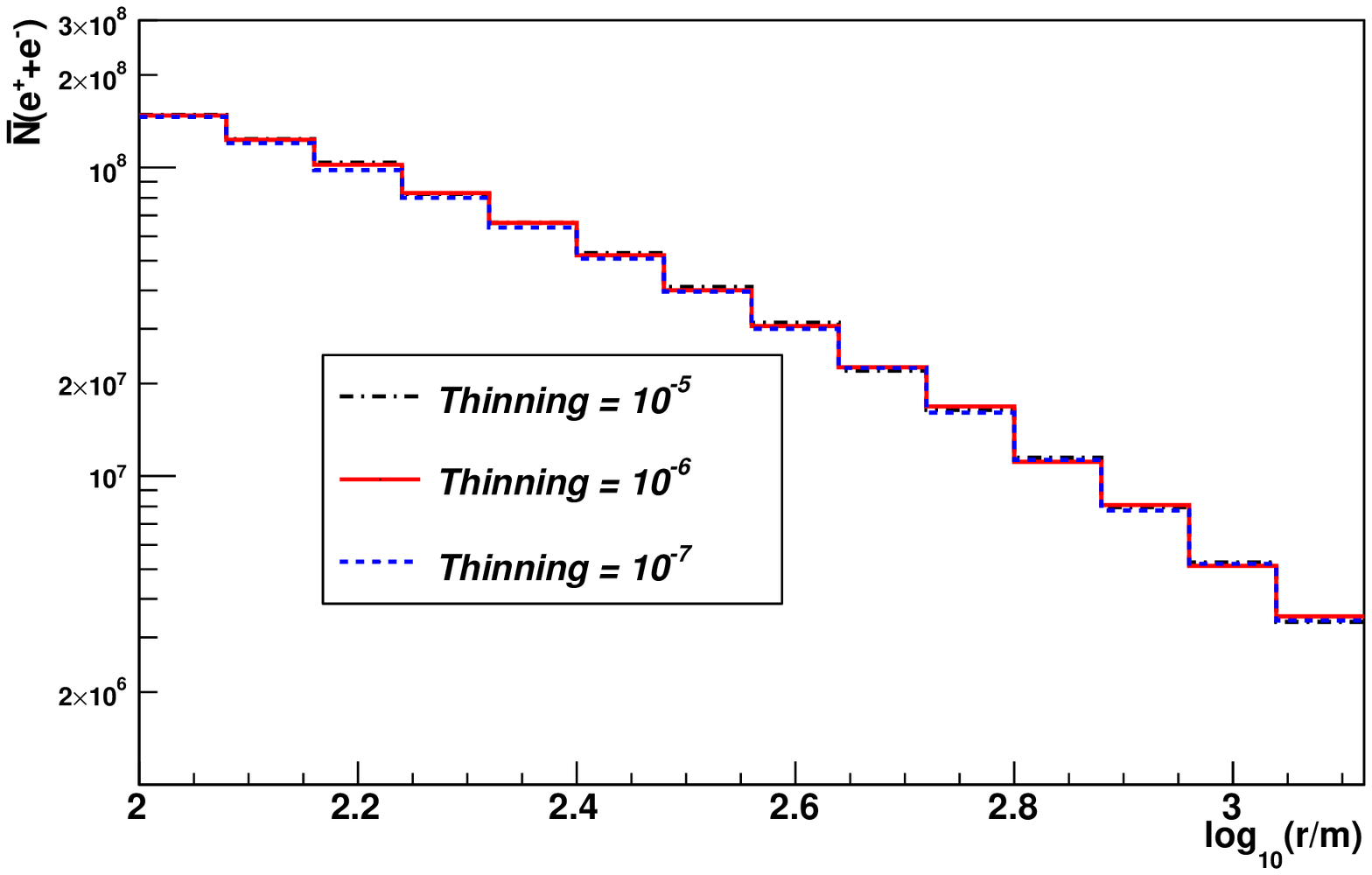}
&\includegraphics[angle=0, width=7.0cm]{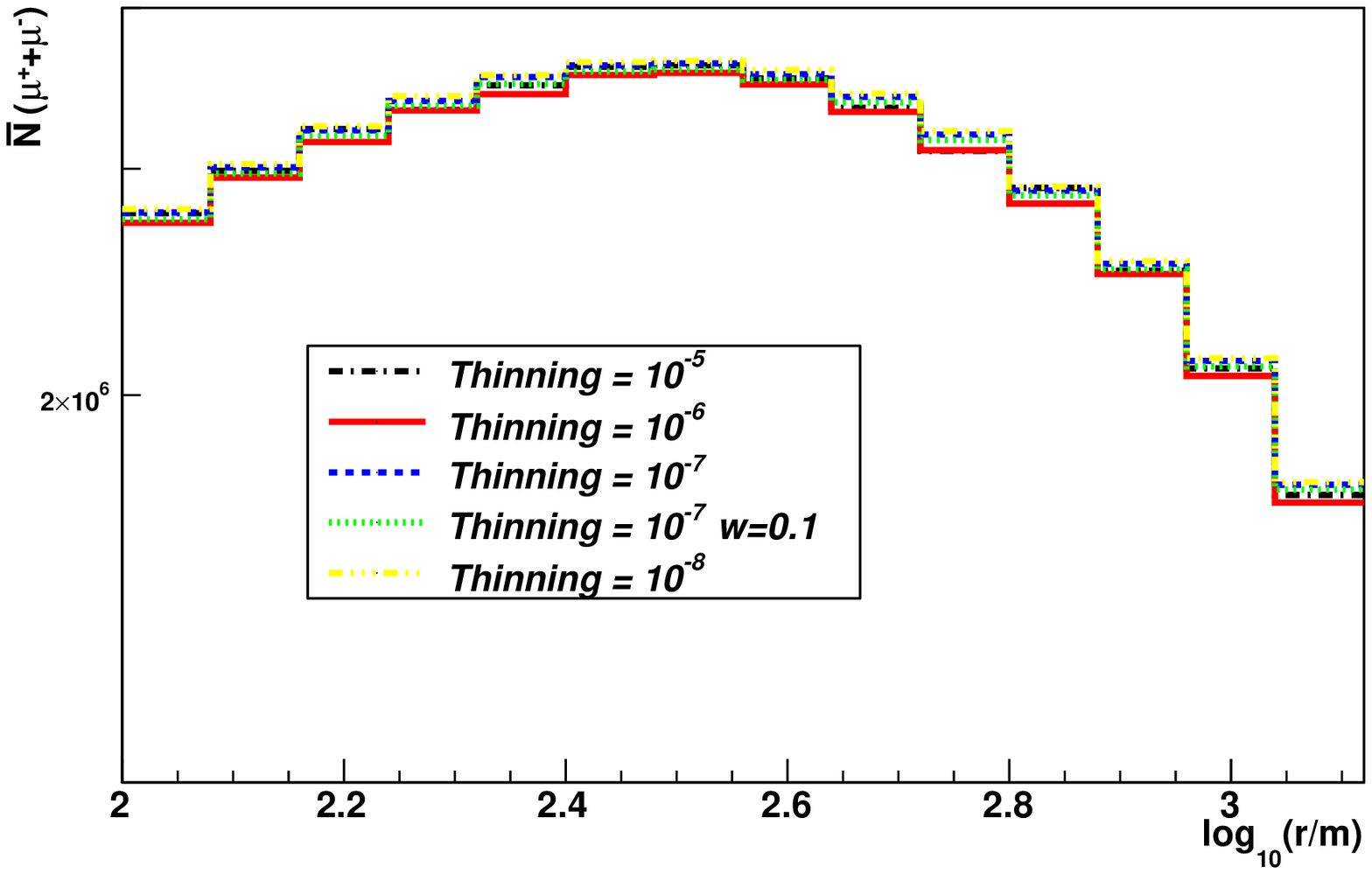}
\\
\end{tabular}\par}
\caption{Average number $\bar N$ of
  electrons (left panel) and muons (right panel) at ground as a
  function of the logarithm of the distance to the core. 100 proton-showers
of energy $10^{19}$ eV and $\theta=0^\circ$ were simulated with different thinning
levels as indicated in the insets. 
}
\label{fig:fig6}
\end{figure}

\begin{figure}[ht]
{\centering \begin{tabular}{cc}
\includegraphics[angle=0, width=7.0cm]{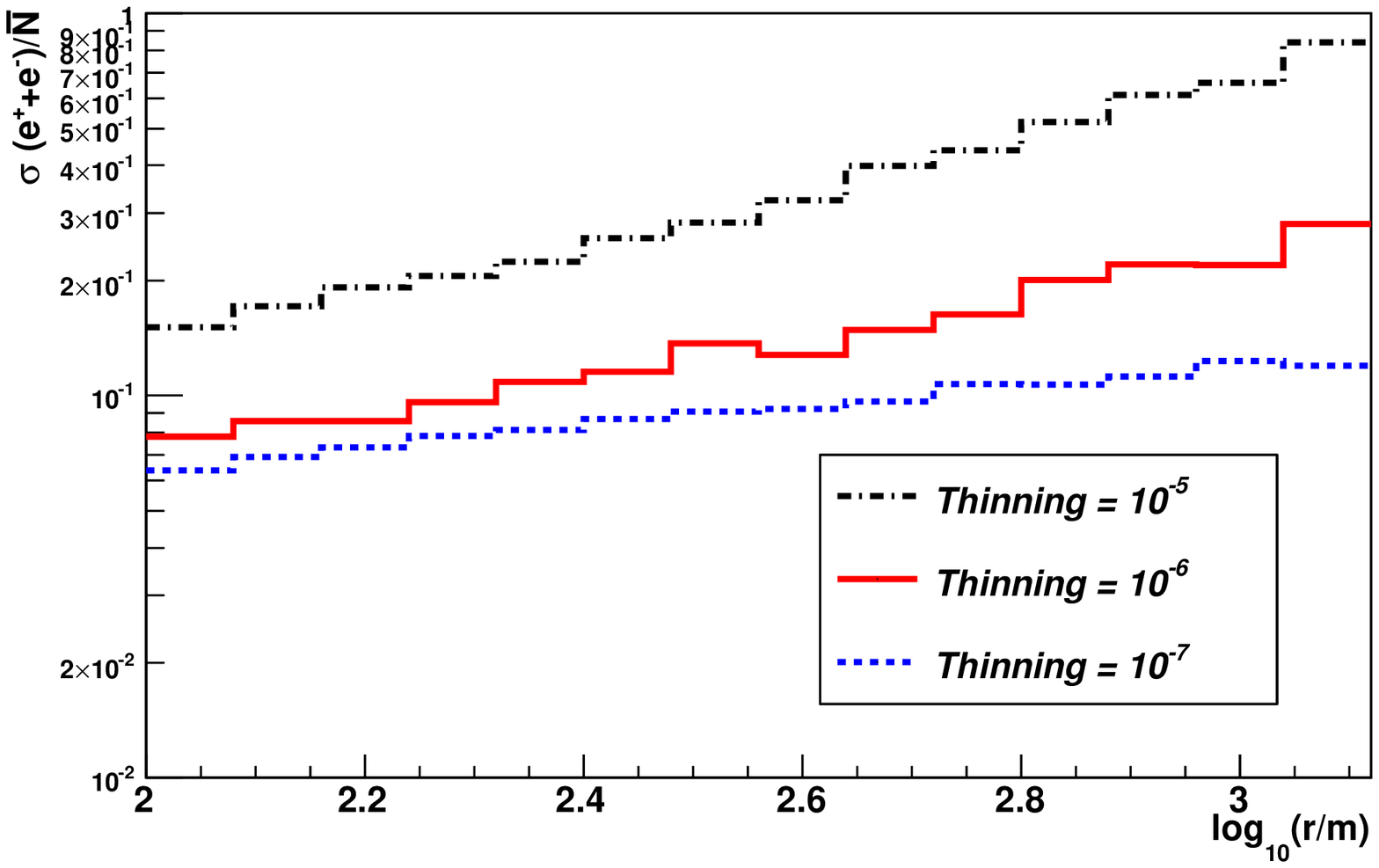}
&\includegraphics[angle=0, width=7.0cm]{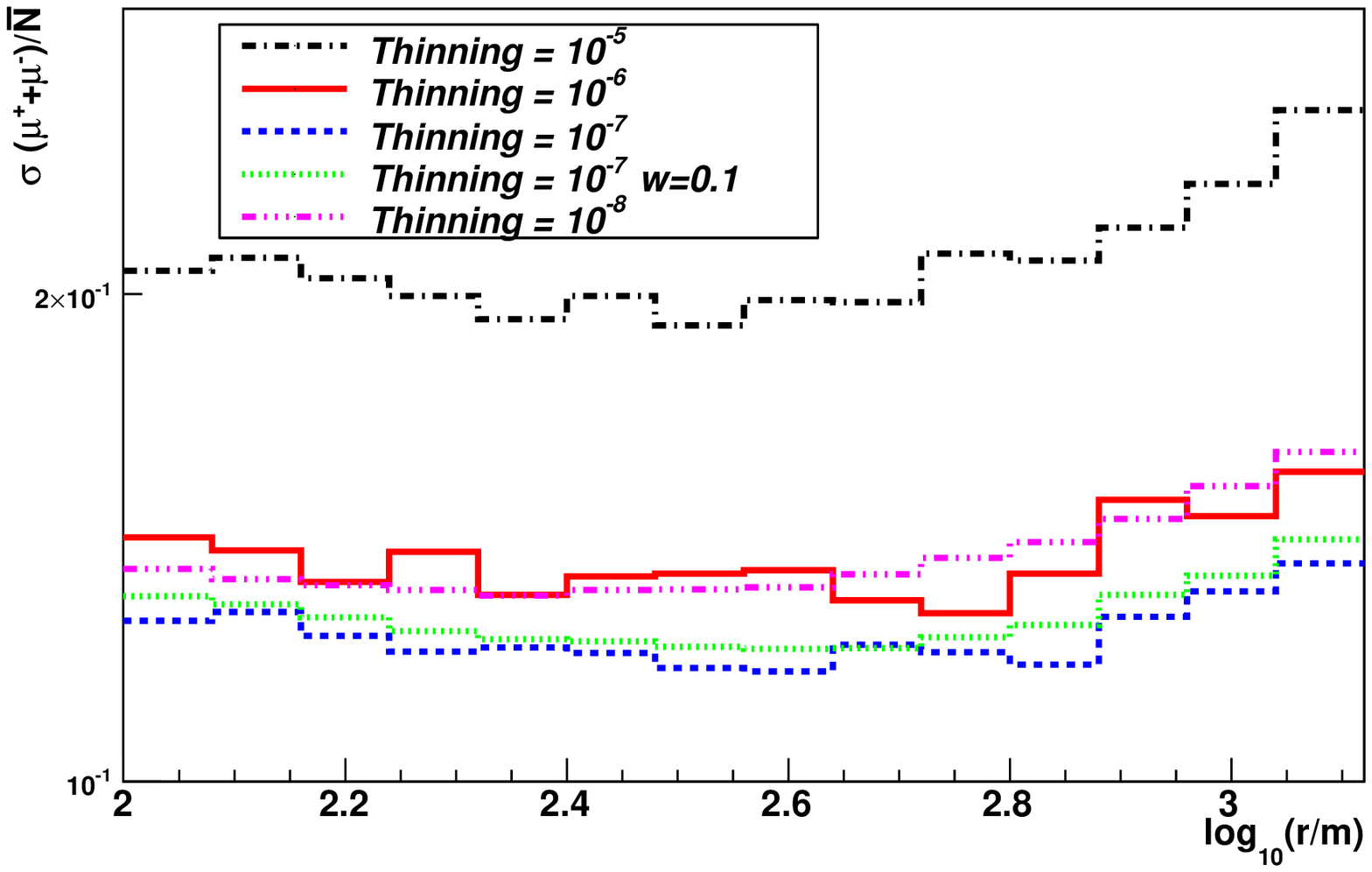}
\\
\end{tabular}\par}
\caption{The relative fluctuation $\sigma/\bar N$ in the number of
  electrons (left panel) and muons (right panel) at ground, as a
  function of the logarithm of the distance to the shower core, for the same
  sets of shower simulations as in Fig.~\ref{fig:fig6}.
}
\label{fig:fig7}
\end{figure}

Firstly in Fig.~\ref{fig:fig6} we show the average number $\bar N$ of electrons 
(left panel) and muons (right panel) versus the distance to the shower axis $r$ 
for different thinning levels. As can be seen, the result is rather independent of the thinning
level. This is not the case for the relative fluctuations $\sigma/{\bar N}$ shown in Fig.~\ref{fig:fig7},
which depend strongly on the thinning level used in the simulations.
For both electron and muons $\sigma/\bar N$ has a
minimum at the distance at which the number of particles is largest.
Also, as expected, the fluctuations can be seen to converge to a common 
value at each distance as the thinning level decreases, because
the effect of thinning is increasingly less important. 
The artificial fluctuations introduced by thinning  
in the case of electrons, do not contribute equally to the total fluctuation 
at all distances from the core as expected. 
It can be seen for instance that the relative fluctuation rises with $r$ 
but the increase is smaller the smaller the thinning level. 

\begin{figure}[ht]
{\centering \begin{tabular}{cc}
\includegraphics[angle=0, width=7.0cm]{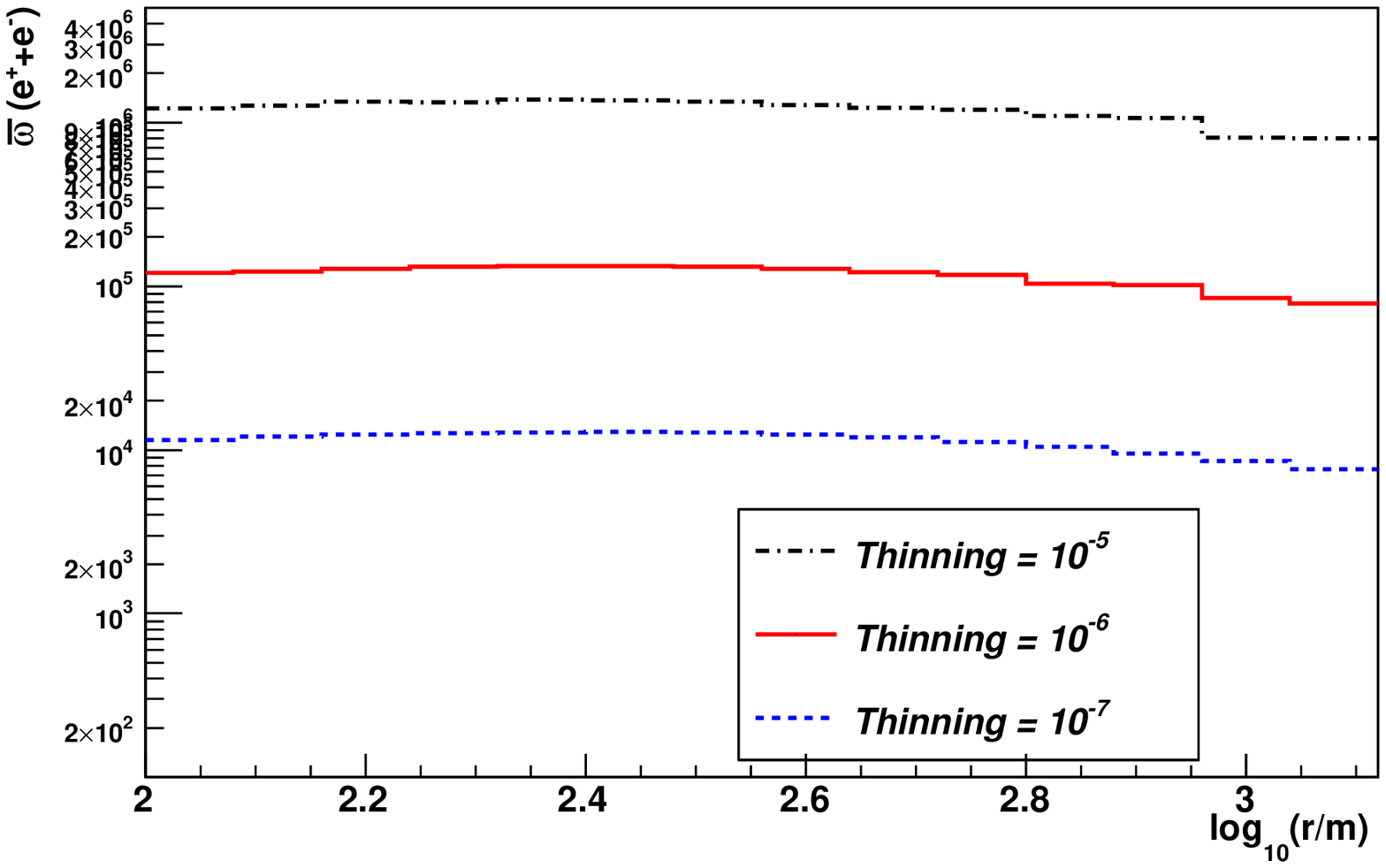}
&\includegraphics[angle=0, width=7.0cm]{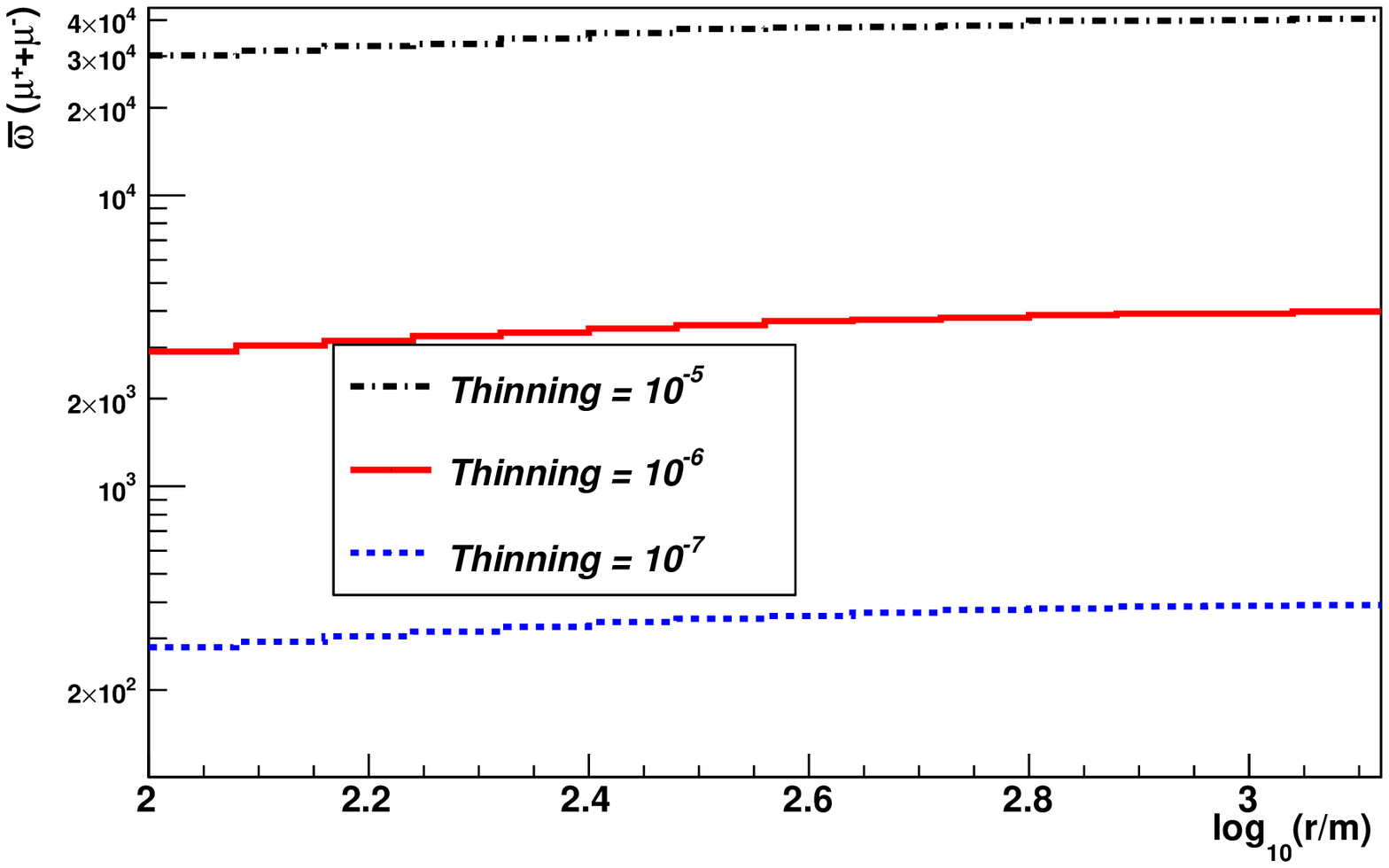}
\\
\end{tabular}\par}
\caption{Average weight $\bar w$ of the distribution of weights 
of electrons (left panel) and muons (right panel) at ground, as 
a function of the logarithm of the distance to the shower core, 
 for sets of 100 proton-induced showers at $10^{19}$ eV and $\theta=0^\circ$, simulated with
different thinning levels as indicated in the insets.
}
\label{fig:fig8}
\end{figure}

In Fig.~\ref{fig:fig8} we plot the average weight $\bar w$ assigned in the 
process of thinning to electrons (left panel) and muons (right panel) as a function of the distance to
the shower axis. In both cases the average weight is simply proportional to the
thinning level $\bar w \propto R_{\rm thin}$, as expected. The average weight
of electrons is typically 50 times larger than that of the muons. For electrons 
the average weight decreases at large distances to the core, because
far from the core the electrons are mainly produced by muon decay, and muons carry
a smaller weight. For the muons, there is a mild increase with $r$
because the highest energy muons, which typically carry a small weight 
(i.e. they are less thinned than lower energy muons), are typically produced close to shower 
axis.  

\begin{figure}[ht]
{\centering \begin{tabular}{cc}
\includegraphics[angle=0, width=7.0cm]{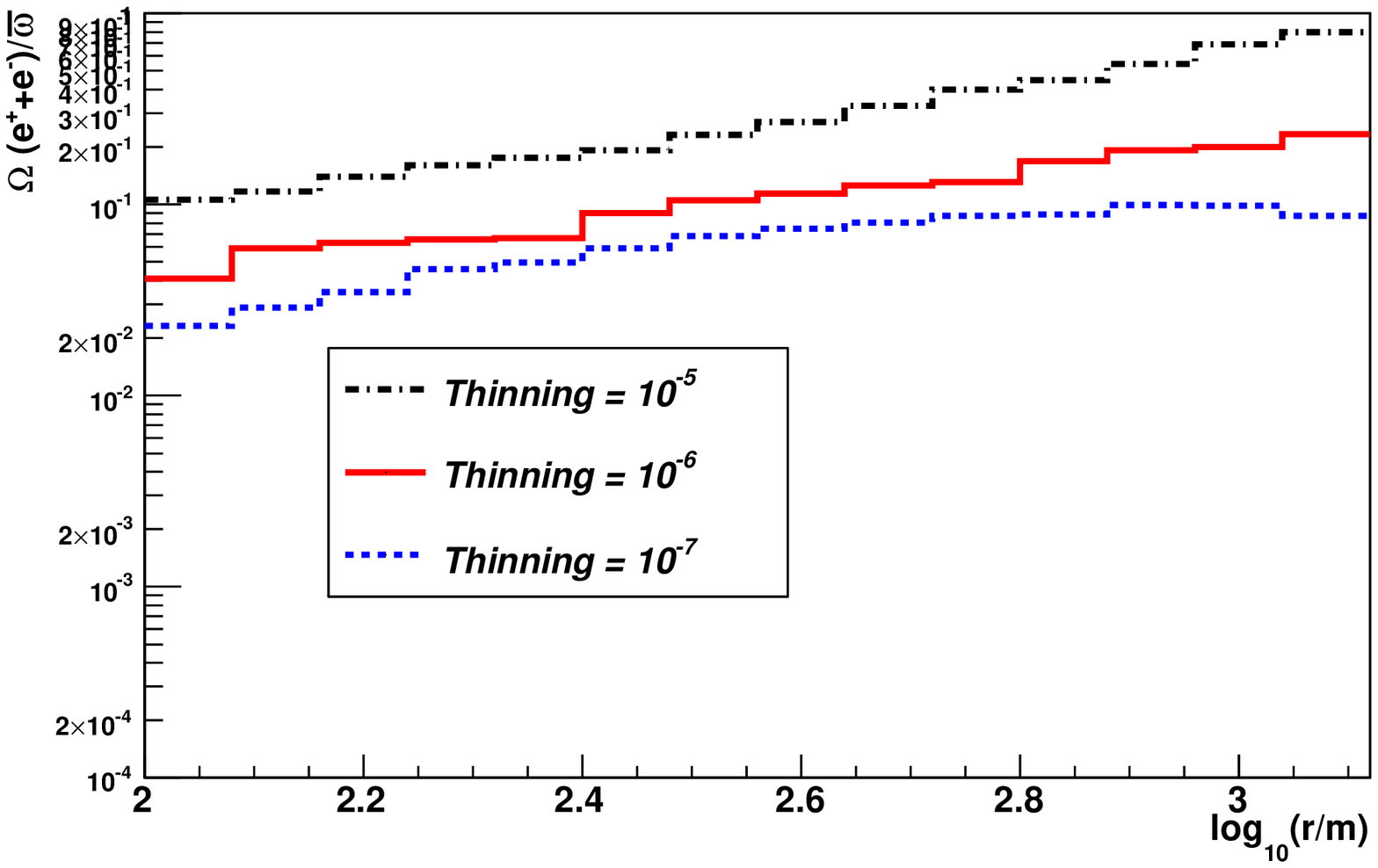}
&\includegraphics[angle=0, width=7.0cm]{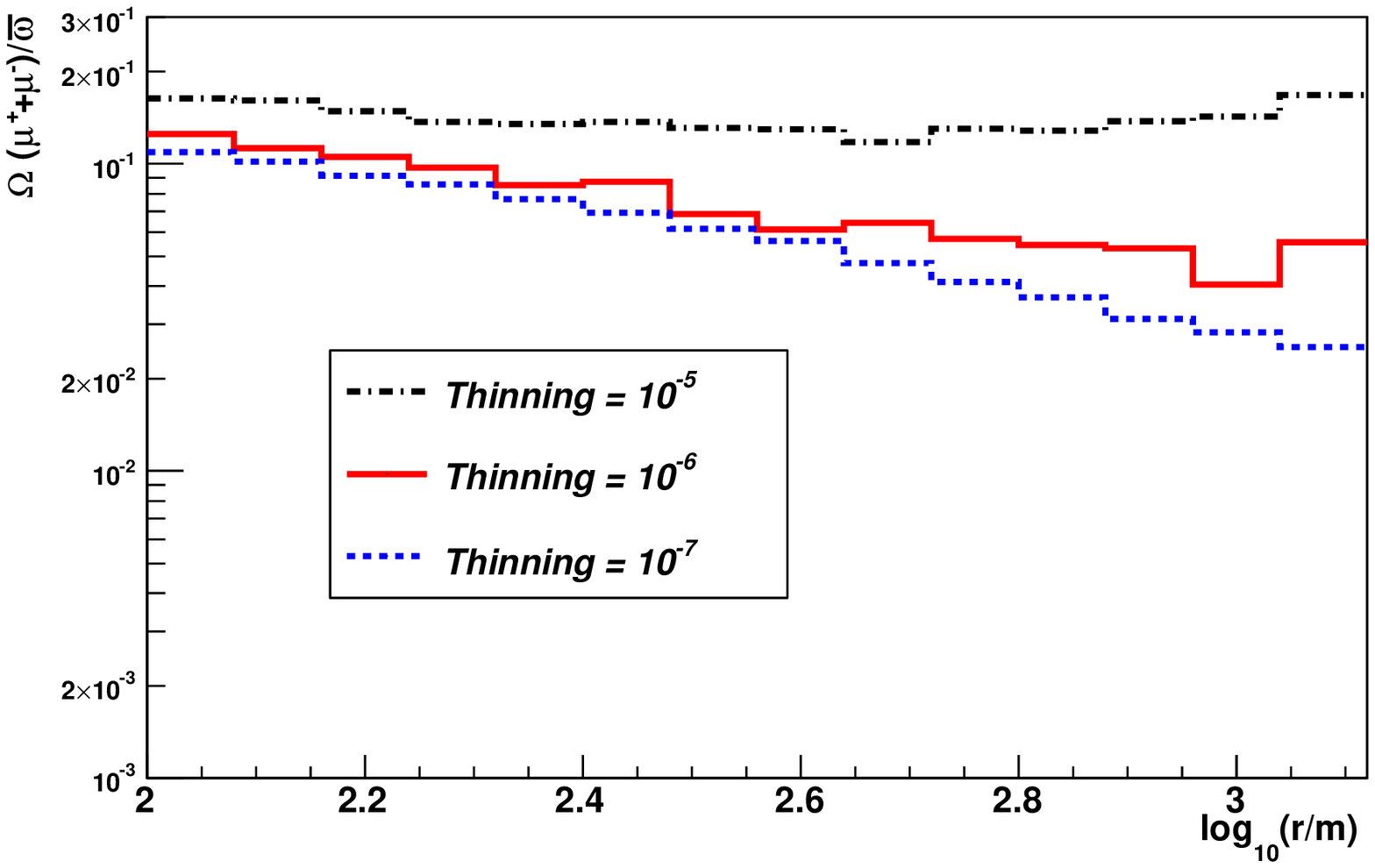}
\\
\end{tabular}\par}
\caption{Relative fluctuations of the distribution of weights ($\Omega/\bar w$) 
of electrons
  (left panel) and muons (right panel) at ground as a function of the
  logarithm of the distance to the core,
for the same sets of shower simulations as in Fig.~\ref{fig:fig8}.
}
\label{fig:Omega}
\end{figure}

In Fig.~\ref{fig:Omega} we plot the relative fluctuations of the 
distribution of weights $\Omega/\bar w$, for electrons (left panel) 
and muons (right panel) as a function of 
distance to the shower core. As expected the fluctuation of the weight
decreases as $R_{\rm thin}$ decreases and the showers are less thinned. 
Also, for values of $R_{\rm thin}<10^{-5}$ the relative
fluctuation $\Omega/\bar w$ is roughly independent of the thinning level, 
and as a consequence we have that approximately $\Omega \propto R_{\rm thin}$. 

\begin{figure}[ht]
{\centering \begin{tabular}{cc}
\includegraphics[angle=0, width=7.0cm]{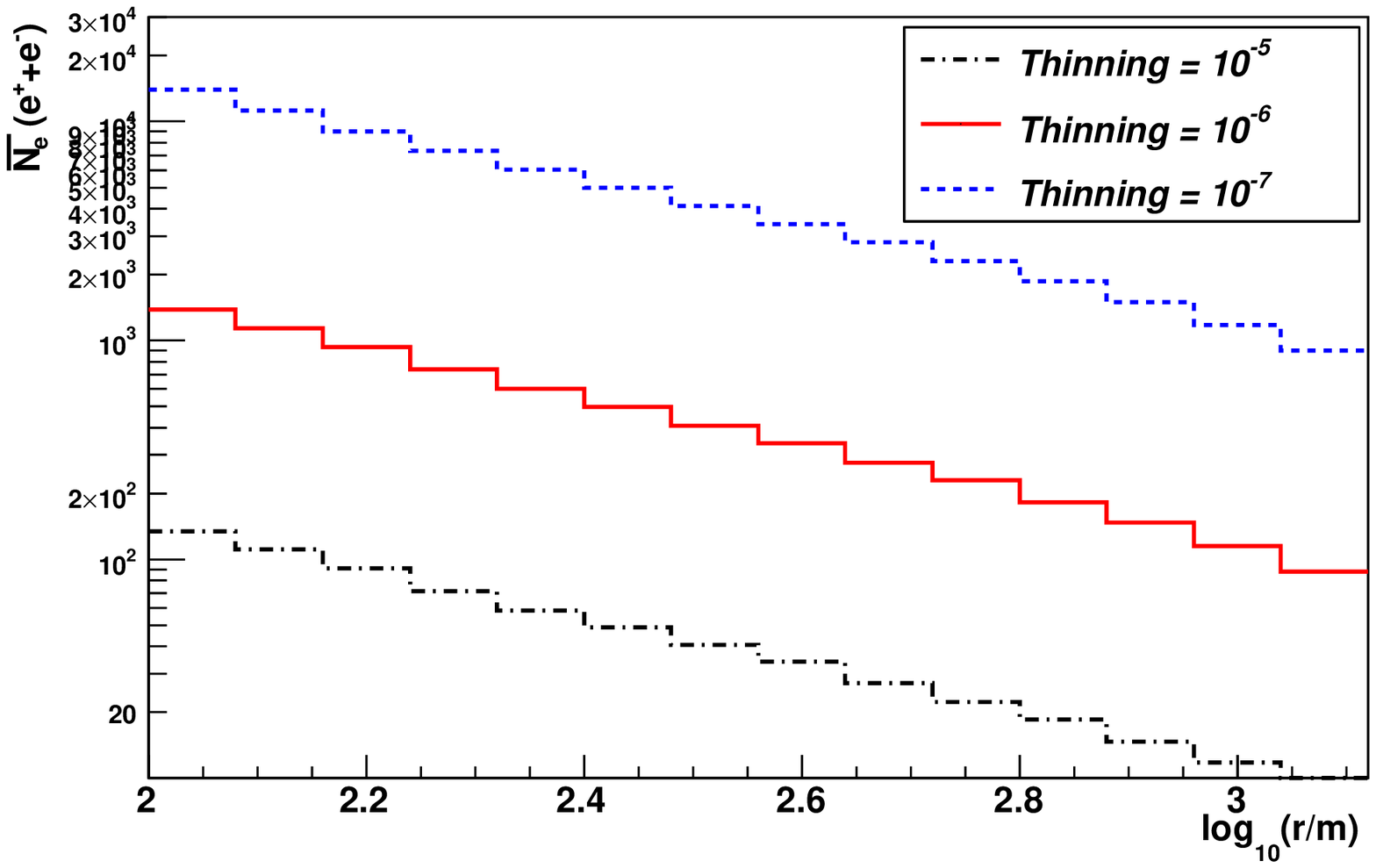}
&\includegraphics[angle=0, width=7.0cm]{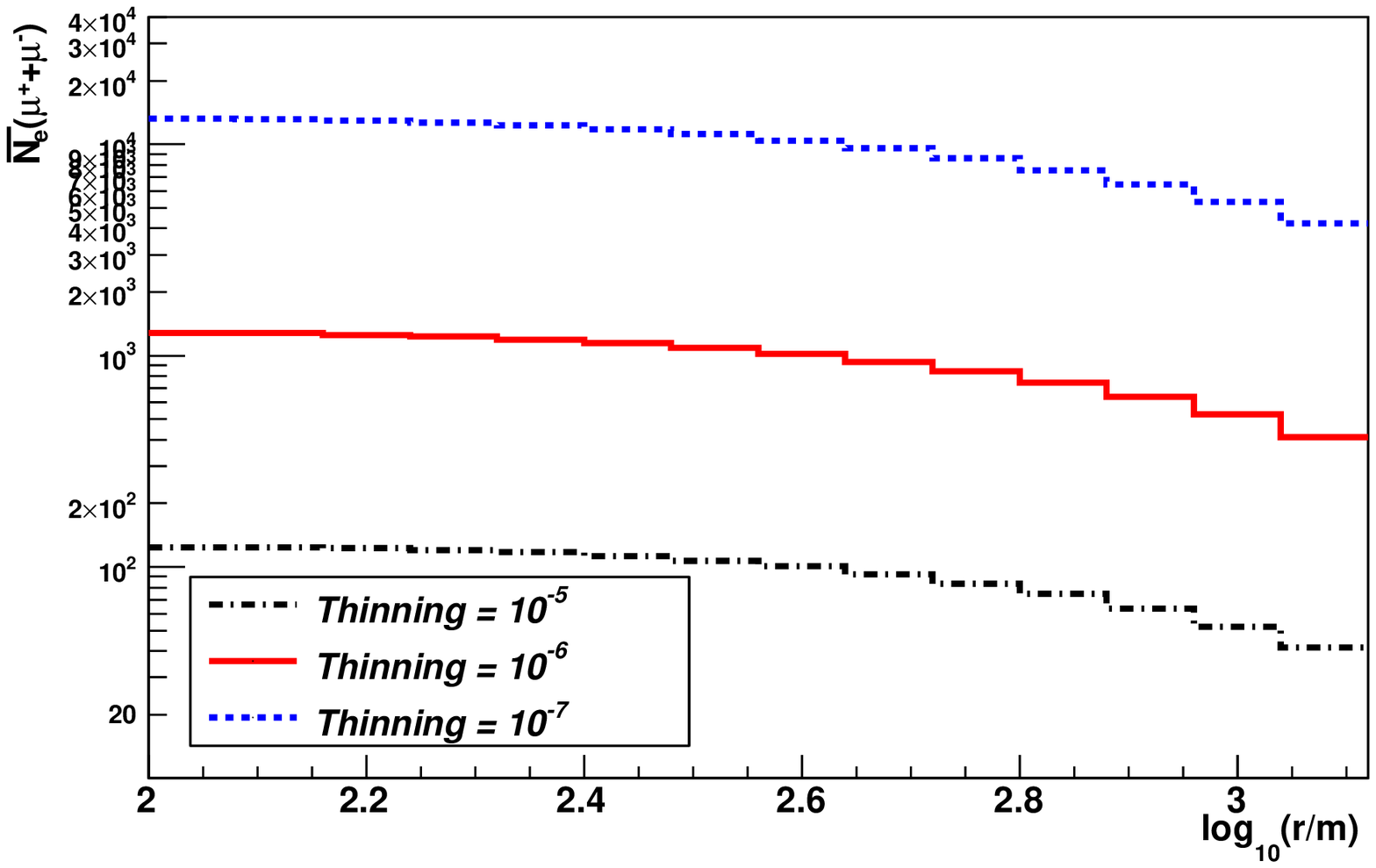}
\\
\end{tabular}\par}
\caption{Average number of entries (non-thinned particles) 
$\bar N_e$ for electrons (left panel) and muons (right panel) at ground, 
as a function of the logarithm of the distance to the core,
for the same sets of shower simulations as in Fig.~\ref{fig:fig8}.
% for 10 EeV proton showers.  The black line corresponds to a
%  relative thinning of 10$^{-5}$, the red line to 10$^{-6}$, and the blue line
%  to 10$^{-7}$.
}
\label{fig:Ne}
\end{figure}

\begin{figure}[ht]
{\centering \begin{tabular}{cc}
\includegraphics[angle=0, width=7.0cm]{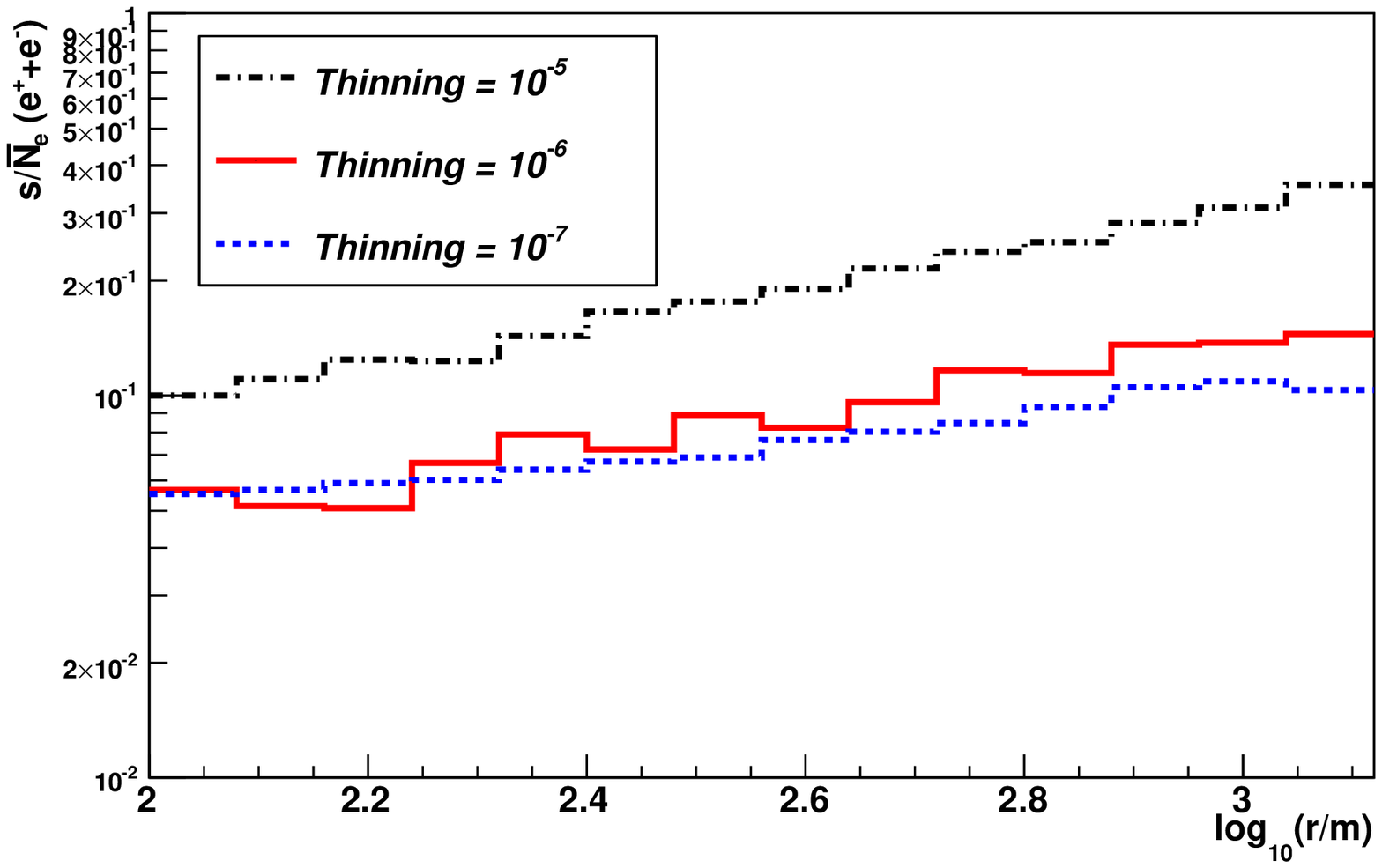}
&\includegraphics[angle=0, width=7.0cm]{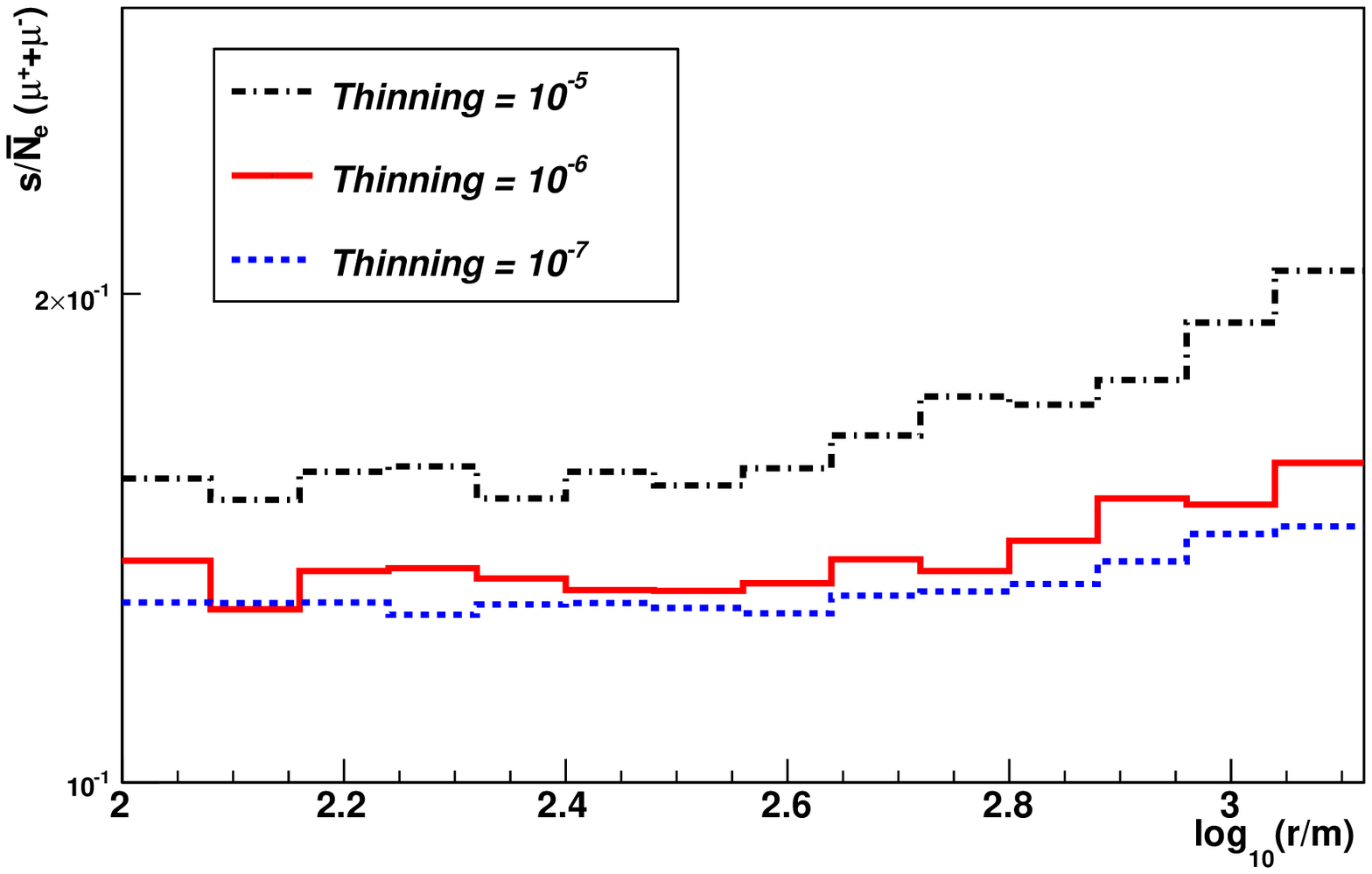}
\\
\end{tabular}\par}
\caption{Relative fluctuation ($s/\bar N_e$) of the distribution of the 
number of entries for electrons (left panel) and muons (right panel) at ground,
 as a function of the logarithm of the distance to the core,
for the same sets of shower simulations as in Fig.~\ref{fig:fig8}.
}
\label{fig:s}
\end{figure}

In Fig.~\ref{fig:Ne}, we show the average number of entries $\bar N_e$ 
for the same simulations as in Fig.~\ref{fig:fig8} above. 
Clearly, we have that $\bar N_e \propto R_{\rm thin}^{-1}$ as imposed by
the constraint in Eq.~(\ref{eq:mean}) that the average number of particles $\bar N$
has to be independent of $R_{\rm thin}$, together with the fact that $\bar w \propto R_{\rm thin}$. 
In Fig.~\ref{fig:s}, we show the relative 
fluctuation in the number of entries $s/{\bar N_e}$. For small thinning levels 
($R_{\rm thin} < 10^{-6}$), we find that the relative fluctuation is approximately 
independent of the thinning level, implying that $s \propto R_{\rm thin}^{-1}$.

\begin{figure}[ht]
{\centering \begin{tabular}{cc}
\includegraphics[angle=0, width=7.0cm]{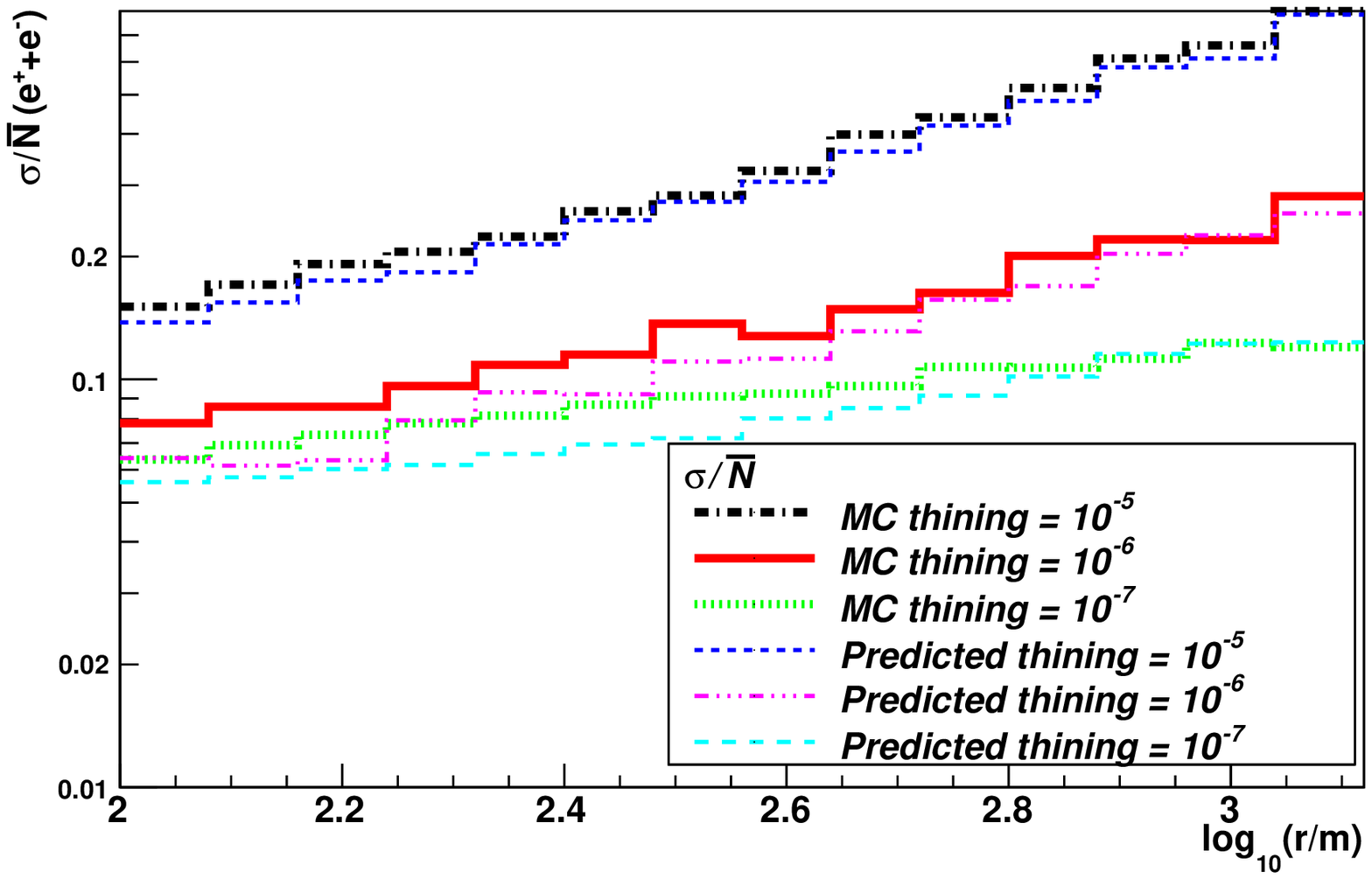}
&\includegraphics[angle=0, width=7.0cm]{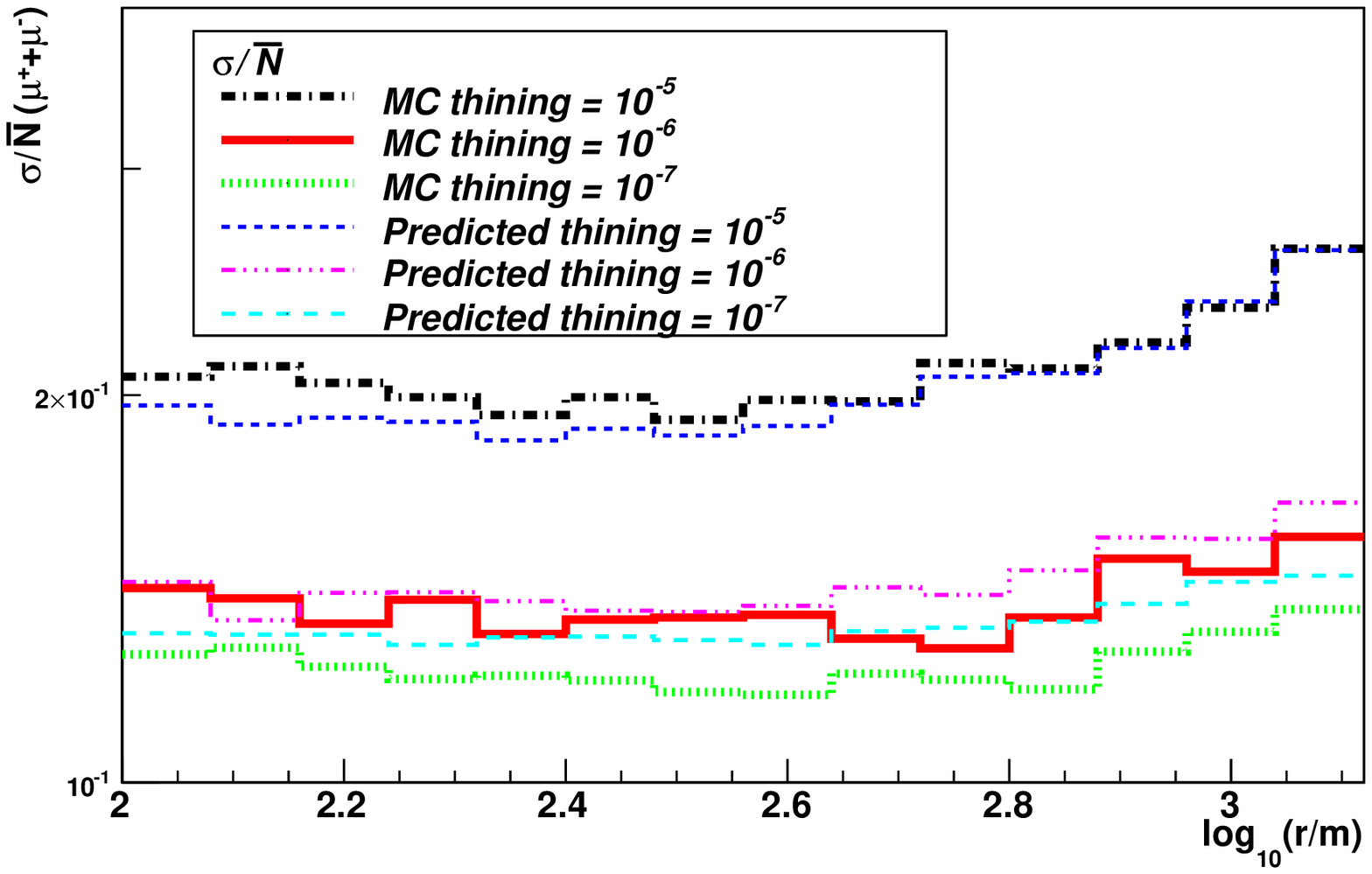}
\\
\end{tabular}\par}
\caption{Relative fluctuations $\sigma/\bar N$ of the distribution of 
number of electron (left panel) and
  muons (right panel) at ground, as a function of the logarithm of the distance
  to the core,
for the same sets of shower simulations as in Fig.~\ref{fig:fig8}.
%  The upper lines correspond to simulations performed with relative
%thinning level $R_{\rm thin}=10^{-5}$ and the lower lines are for $R_{\rm thin}=10^{-7}$. 
  The $\sigma/\bar N$ obtained in the simulations (MC) is compared to that predicted by 
Eq.~(\ref{eq:sigma}).}
\label{fig:formula}
\end{figure}

Finally, in Fig.~\ref{fig:formula}, we compare the relative fluctuation
of the number of particles $\sigma/\bar N$ obtained directly in Monte Carlo simulations, 
with that predicted by Eq.~(\ref{eq:sigma}), using the values of $\bar w$, $\Omega$,
$\bar N_e$ and $s$ obtained in the same simulations. The 
comparison is shown for thinning levels $R_{\rm thin}=10^{-5},~10^{-6}$ and $10^{-7}$. 
The agreement between the  $\sigma/\bar N$ obtained in Monte Carlo simulations, 
and that predicted by Eq.~(\ref{eq:sigma}) is at the level of $<20\%$ for electrons
and $<5\%$ for muons, confirming that  
Eq.~(\ref{eq:sigma}) accounts for all the fluctuations
(artificial and physical) appearing in the simulations of EAS with thinning.

\begin{figure}[ht]
{\centering \begin{tabular}{cc}
\includegraphics[angle=0, width=7.0cm]{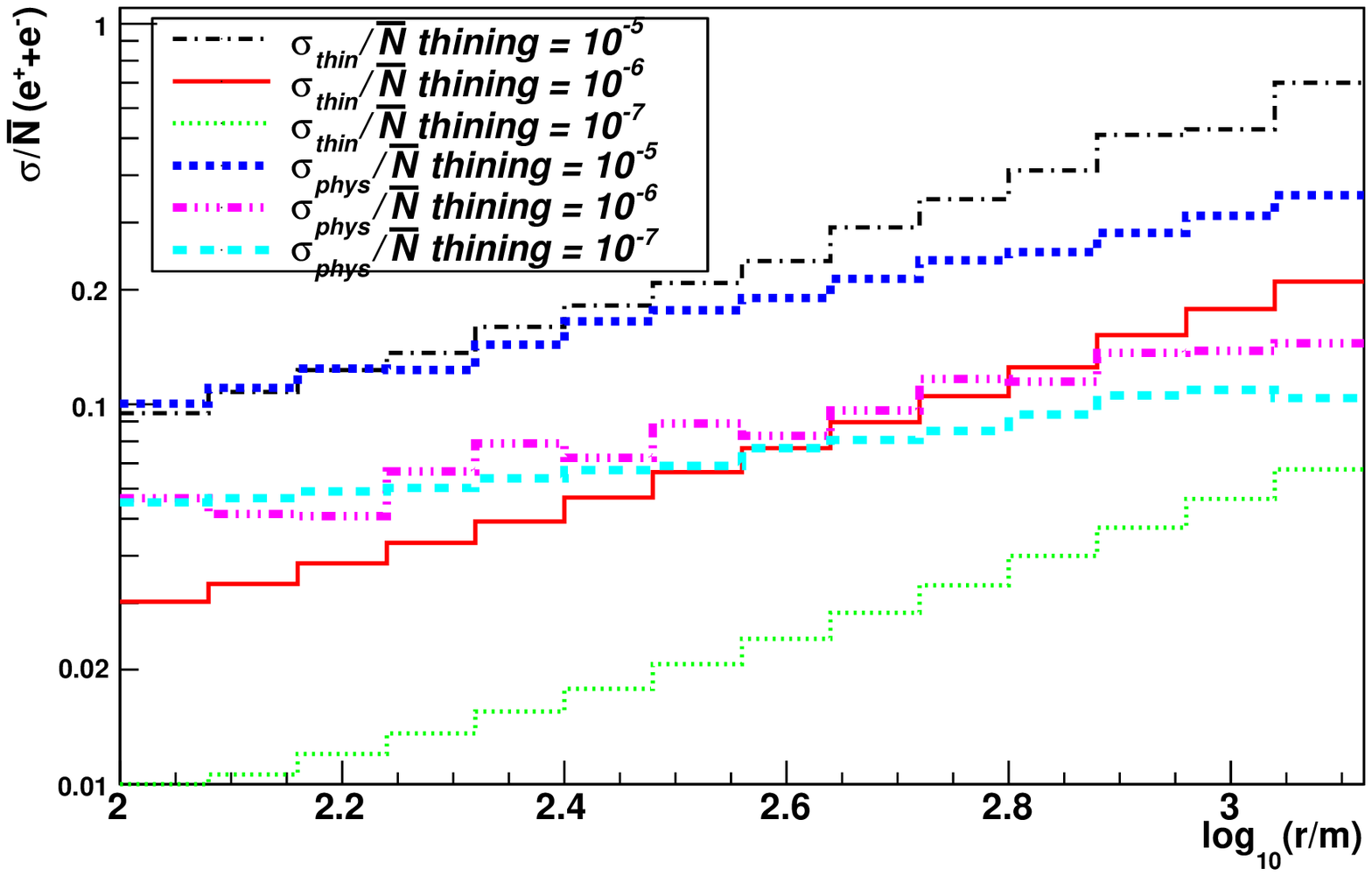}
&\includegraphics[angle=0, width=7.0cm]{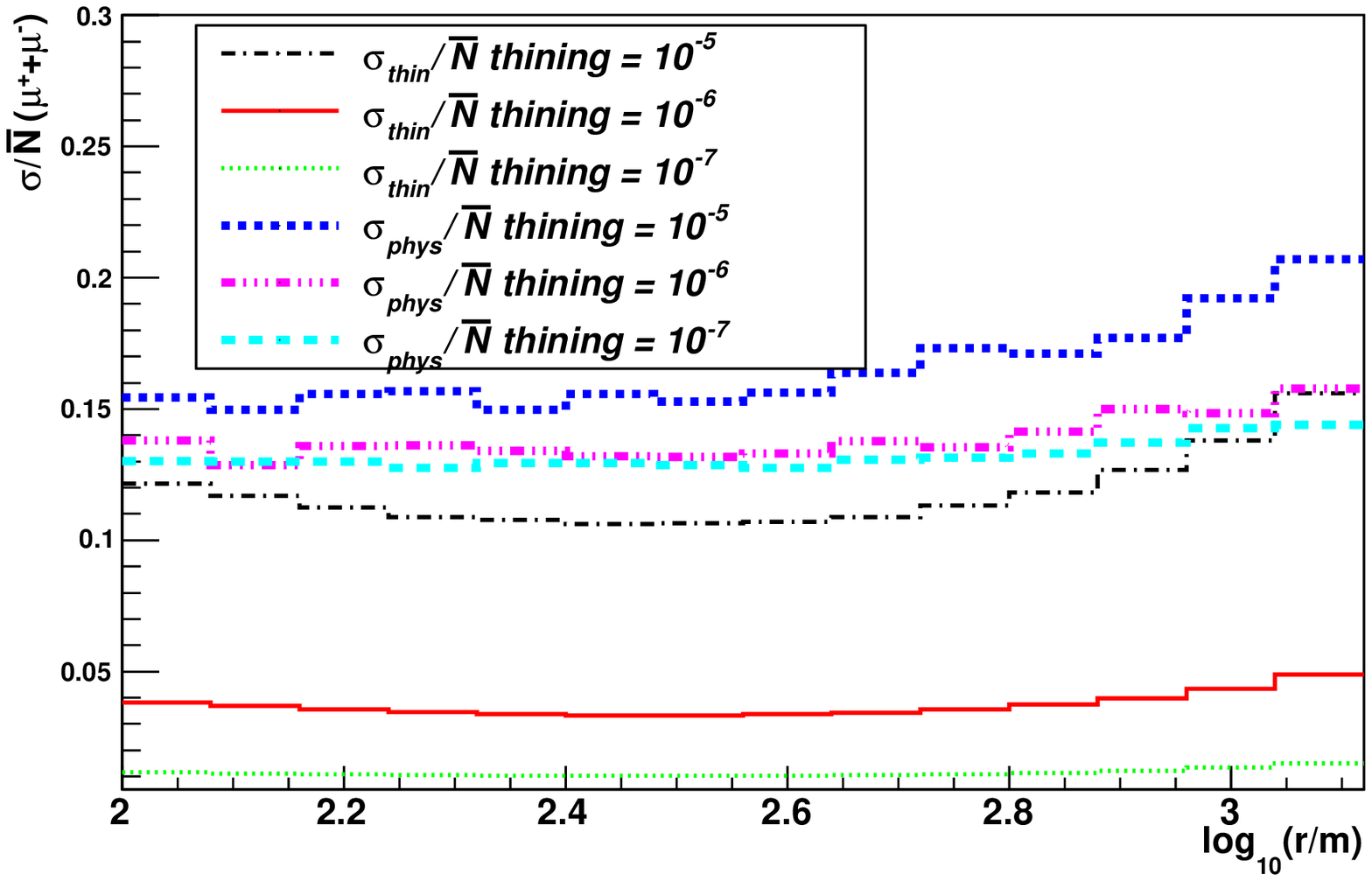}
\\
\end{tabular}\par}
\caption{Relative physical $\sigma_{\rm phys}/\bar N$ 
and thinning fluctuations $\sigma_{\rm thin}/\bar N$ (as predicted
by Eq.~(\ref{eq:sigma})) of the distribution of 
number of electron (left panel) and
  muons (right panel) at ground, as a function of the logarithm of the distance
  to the core,
for the same sets of shower simulations as in Fig.~\ref{fig:fig8}.
}
\label{fig:formula2}
\end{figure}

From the scalings with $R_{\rm thin}$ of the different 
magnitudes involved in Eq.~(\ref{eq:sigma}) obtained before, it is straightforward to deduce that  
$\sigma_{\rm thin} \propto R_{\rm thin}$, while $\sigma_{\rm phys}$  
should be approximately independent of $R_{\rm thin}$. This is seen in 
Fig.~\ref{fig:formula2}: 
For thinning levels $R_{\rm thin}\sim 10^{-6}$ and smaller,
$\sigma_{\rm phys}$ is almost independent of $R_{\rm thin}$,
while $\sigma_{\rm thin}$ depends strongly on $R_{\rm thin}$.
In Fig.~\ref{fig:formula2}, it can also be seen that
as the thinning level decreases the $\sigma_{\rm phys}$ term 
increasingly dominates. 
This is of course expected, but it is remarkable that it was
obtained from Monte Carlo simulations, and therefore it gives a
strong support to our identification of $\sigma_{\rm phys}$ with the true physical
fluctuations.

\subsubsection{Dependence of fluctuations at ground on the number of particles}

Let us now consider the dependence of the fluctuations 
on the size of the ring 
around a distance to the shower core $r$, where particles are collected in the simulation.
Let us consider how the density of particles is evaluated.  
If $\bar N(r,\Delta r)$ is the average number of particles at distance $r$ in a small
bin $\Delta r$ (here we assume cylindrical symmetry around the shower axis,
but the argument does not depend on this simplification), then the density of
particles $\rho(r)$ can be defined as
\begin{equation}
\rho(r) = {\rm lim}_{\Delta r\rightarrow 0} \; \; \frac{\bar N(r,\Delta r)}{2 \pi r
  \Delta r}. 
\end{equation}
In the limit $\Delta r\rightarrow 0$, 
$\rho(r)$ is finite, at least for $r\neq0$.
However, the same is not true for the fluctuations, so that in general 
one can not define a ``density of fluctuations'' $\rho_\sigma (r)$. We can see this in a simple
example. Assume that $\sigma(r,\Delta r)$ is the standard deviation of the distribution of the number of
particles in a bin of size $\Delta r$ and at a distance $r$. One could
try to define the density of fluctuations as
\begin{equation}
\rho_\sigma(r) = {\rm lim}_{\Delta r\rightarrow 0} \; \; \frac{\sigma(r,\Delta r)}{2 \pi r  \Delta r}. 
\label{lim}
\end{equation}

If the fluctuations in the number of particles at ground were purely Poissonian, 
we would have
\begin{equation}
\sigma(r,\Delta r) = \sqrt{N(r,\Delta r)},
\label{eq:Poissonian}
\end{equation}
and therefore,
\begin{equation}
\rho_\sigma(r) = {\rm lim}_{\Delta r\rightarrow 0} \; \; \frac{\sqrt{N(r,\Delta
    r)} }{2 \pi r  \Delta r} =   {\rm lim}_{\Delta r\rightarrow 0} 
\; \; \frac{\sqrt{2 \pi r \Delta r \rho(r)}}{2 \pi r \Delta r} = 
{\rm lim}_{\Delta r\rightarrow 0} 
\; \; \frac{\sqrt{ \rho(r)}}{ \sqrt{2 \pi r \Delta r}} \rightarrow \infty,
\label{eq:sigma_deltar}
\end{equation}
i.e., we can not define a density of fluctuations for  
Poissonian shower to shower fluctuations in the number of particles. 
The dependence on bin-size has been identified with the fractal
structure of showers \cite{Kiel}. In the case of Poissonian fluctuations in the 
number of particles, the deduced behaviour is 
$\rho_\sigma(r) \propto \Delta r^\alpha$ with $\alpha=-1/2$, and we would be tempted 
to identify the coefficient $\alpha$ with a fractal exponent. However notice that
for a Poissonian process no fractal structure is implied at all and it would
be erroneous to call it a fractal exponent.

On the other hand, in the case in which the fluctuations behave as:
\begin{equation}
\sigma(r,\Delta r) = f(r) \Delta r + {O}(\Delta r^2),
\end{equation}
where $f(r)$ is a function that does not depend on $\Delta r$, one can define
a density of fluctuations as can be shown trivially applying Eq.~(\ref{lim}). Remarkably, this is precisely 
the case of Furry's fluctuations \cite{furry}, in which $\sigma \sim N = 2\pi r \rho(r) \Delta r$
and then $\rho_\sigma$ would be independent of $\Delta r$, i.e. $\alpha=0$.
Recall that Furry statistics appears as a extremely simplified model of
shower fluctuations \cite{furry}, but it does take into account the
branching structure of the shower (and therefore has an implicit
fractal structure included).

To our knowledge, the actual behaviour of the fluctuations of showers and its
dependence with the bin size is an open theoretical problem, with the
theoretical prejudice ranging between two extremes: purely Poissonian fluctuations 
$\sigma \sim {\sqrt N}$, and stronger fluctuations $\sigma \sim N$. 
For instance, for the longitudinal development of showers one can show that
in fact both types of behaviour occur \cite{fluctu}, i.e.,
\begin{equation}
\sigma^2  = a \, N^2 + b \, N,
\end{equation}
where $a$ and $b$ vary slowly with primary
energy \cite{fluctu}. Near the maximum of the shower, the first term dominates and
fluctuations are not Poissonian. For the lateral distribution no such result
exists but one would expect a similar conclusion.

\begin{figure}[ht]
{\centering \begin{tabular}{cc}
\includegraphics[angle=0, width=7.0cm]{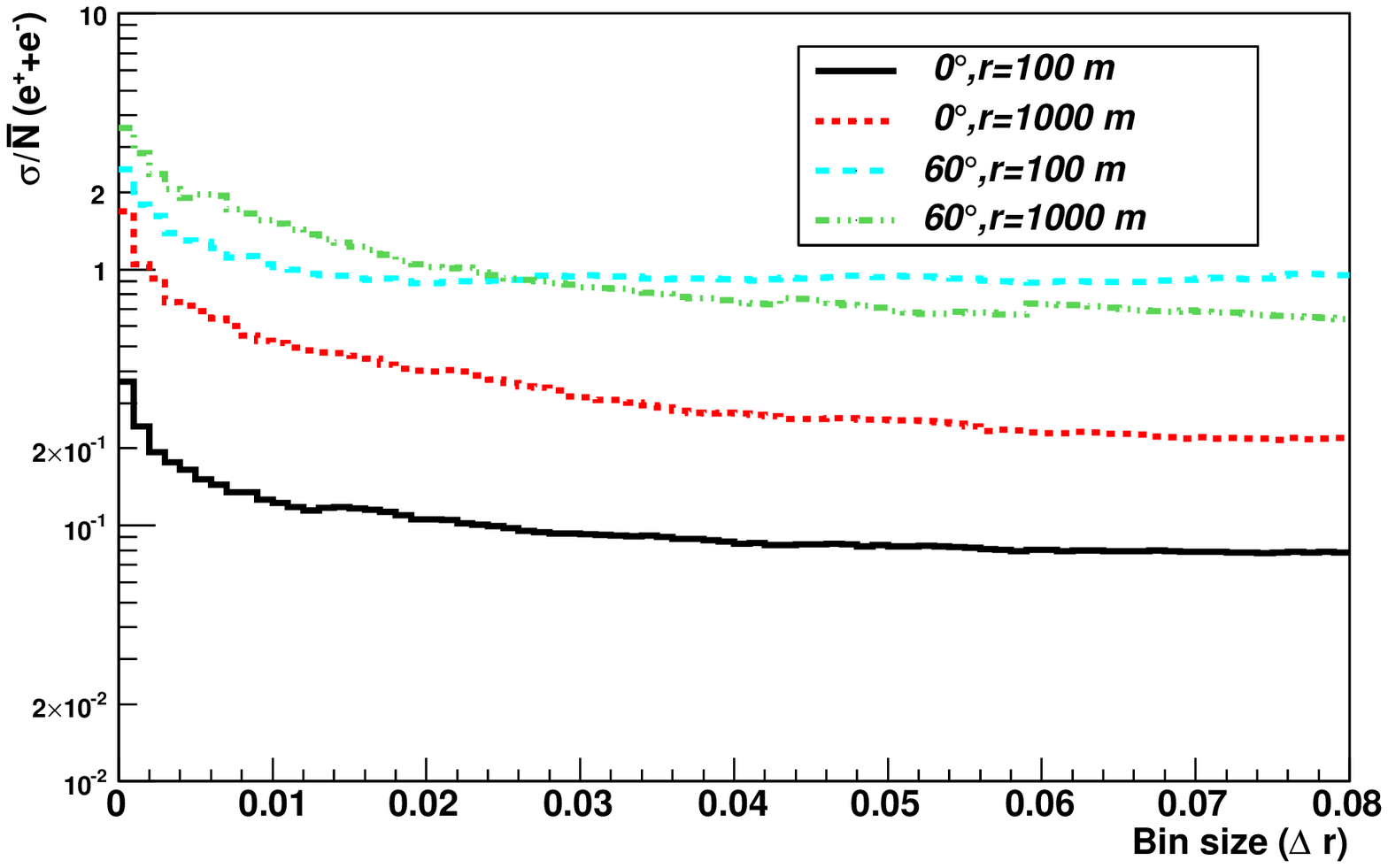}
&\includegraphics[angle=0, width=7.0cm]{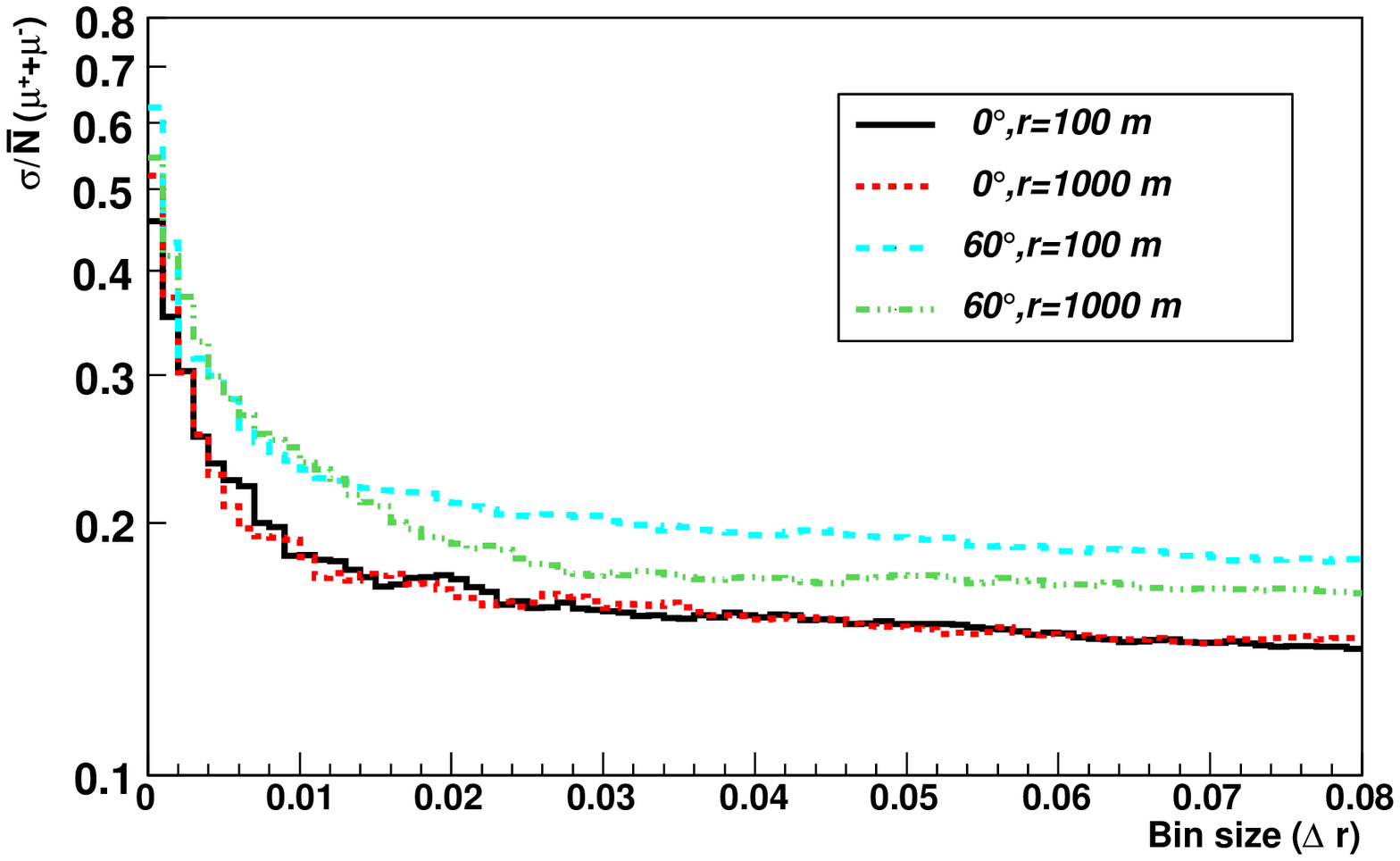}
\\
\end{tabular}\par}
\caption{Relative 
fluctuations $\sigma/\bar N$ of the distribution of 
the number of electrons (left panel) and
  muons (right panel) in a ring of width $\Delta r$, centered 
at different $r$ at ground, as a function of the logarithm of the size of the bin $\Delta r$. 
100 proton-induced showers of $10^{19}$ eV energy were simulated at 
different $\theta$ and fixed thinning level $R_{\rm thin}=10^{-6}$}
\label{fig:bin}
\end{figure}

\begin{figure}[ht]
{\centering \begin{tabular}{cc}
\includegraphics[angle=0, width=7.0cm]{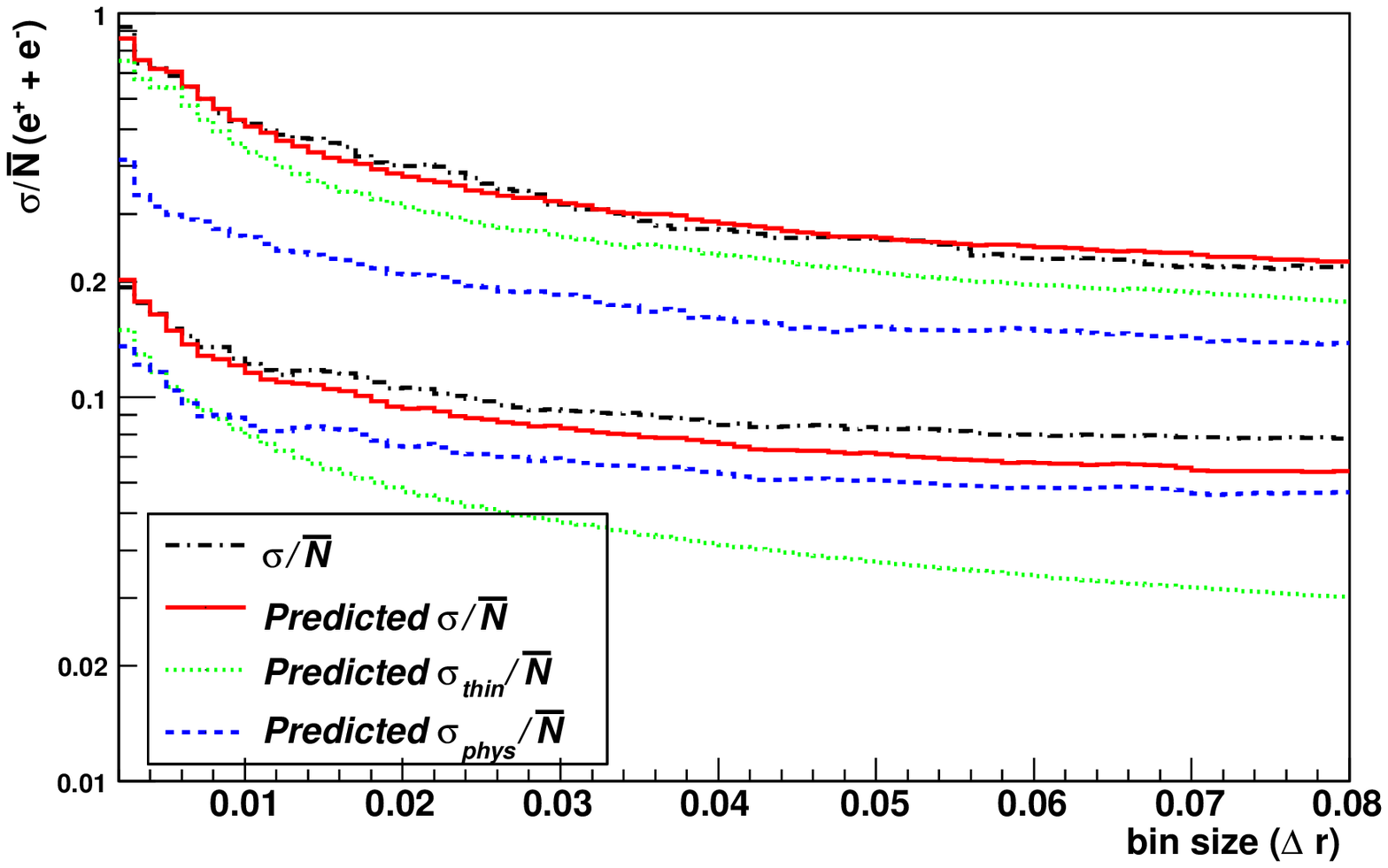}
&\includegraphics[angle=0, width=7.0cm]{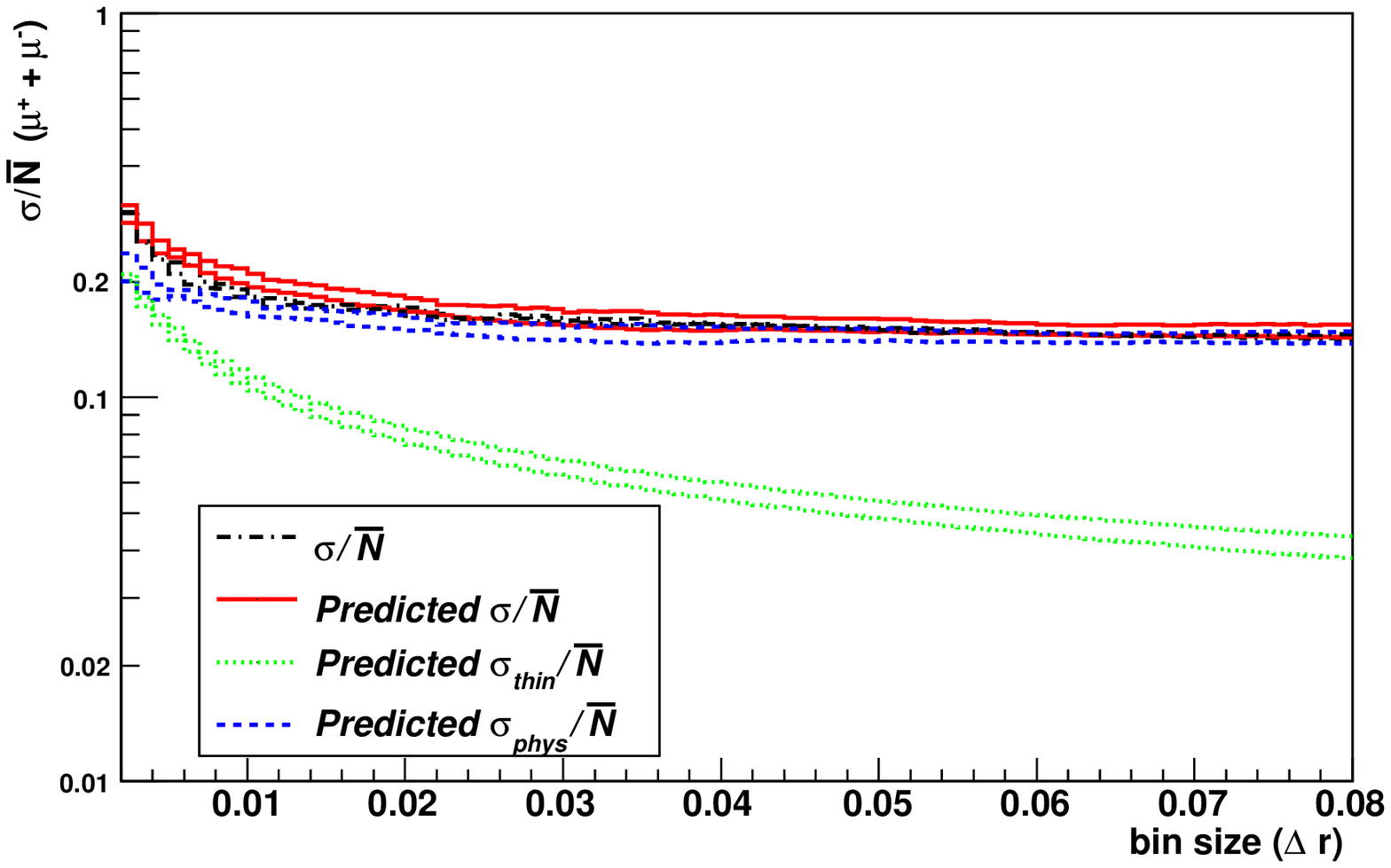}
\\
\end{tabular}\par}
\caption{
Relative fluctuations $\sigma/\bar N$ of the distribution of 
the number of electrons (left panel) and
  muons (right panel) in a ring of width $\Delta r$, centered 
at different $r$ at ground, as a function of the logarithm of size of the bin $\Delta r$.
  The upper lines correspond to simulations performed with relative
thinning level $R_{\rm thin}=10^{-5}$ and the lower lines are for $R_{\rm thin}=10^{-7}$. 
  The $\sigma/\bar N$ obtained in the simulations is compared to that predicted by 
Eq.~(\ref{eq:sigma}). The two terms in Eq.~(\ref{eq:sigma}) corresponding to fluctuations
induced by thinning $\sigma_{\rm thin}$ and physical fluctutations $\sigma_{\rm phys}$ are 
also shown, see insets. 
  } 
\label{fig:bin2}
\end{figure}

In Fig.~\ref{fig:bin}, we show
the relative fluctuation in the number of particles $\sigma/\bar N$,
as a function of the bin size $\Delta r$, for
several zenith angles $\theta$ and distances $r$ to the shower axis. 
Note that $\sigma/\bar N$ is equal to $\rho_\sigma$ in the limit of 
$\Delta r\rightarrow 0$. A fit to a power law
dependence (as suggested by the discussion above) gives
$\sigma/\bar N \propto \Delta r^\alpha$ with $\alpha = -1/2$ for both 
electrons and muons. As explained above this is suggestive of the
fluctuations being Poissonian. However, 
if we split the fluctuations with the aid of Eq.~(\ref{eq:sigma}) into
thinning fluctuations, and physical fluctuations, it can be seen in
Fig.~\ref{fig:bin2} that $\sigma_{\rm phys}/\bar N$ is consistent with being flat
with $\Delta r$, whereas 
$\sigma_{\rm thin}/\bar N$ behaves as a power law (a fit gives $\sigma_{\rm thin}/\bar N \propto \Delta r^{-1/2}$). 
These results suggest that the artificial fluctuations are Poissonian, while 
physical fluctuations behave as $\sigma_{\rm phys} \propto N$. 

\begin{figure}[ht]
{\centering \begin{tabular}{cc}
\includegraphics[angle=0, width=7.0cm]{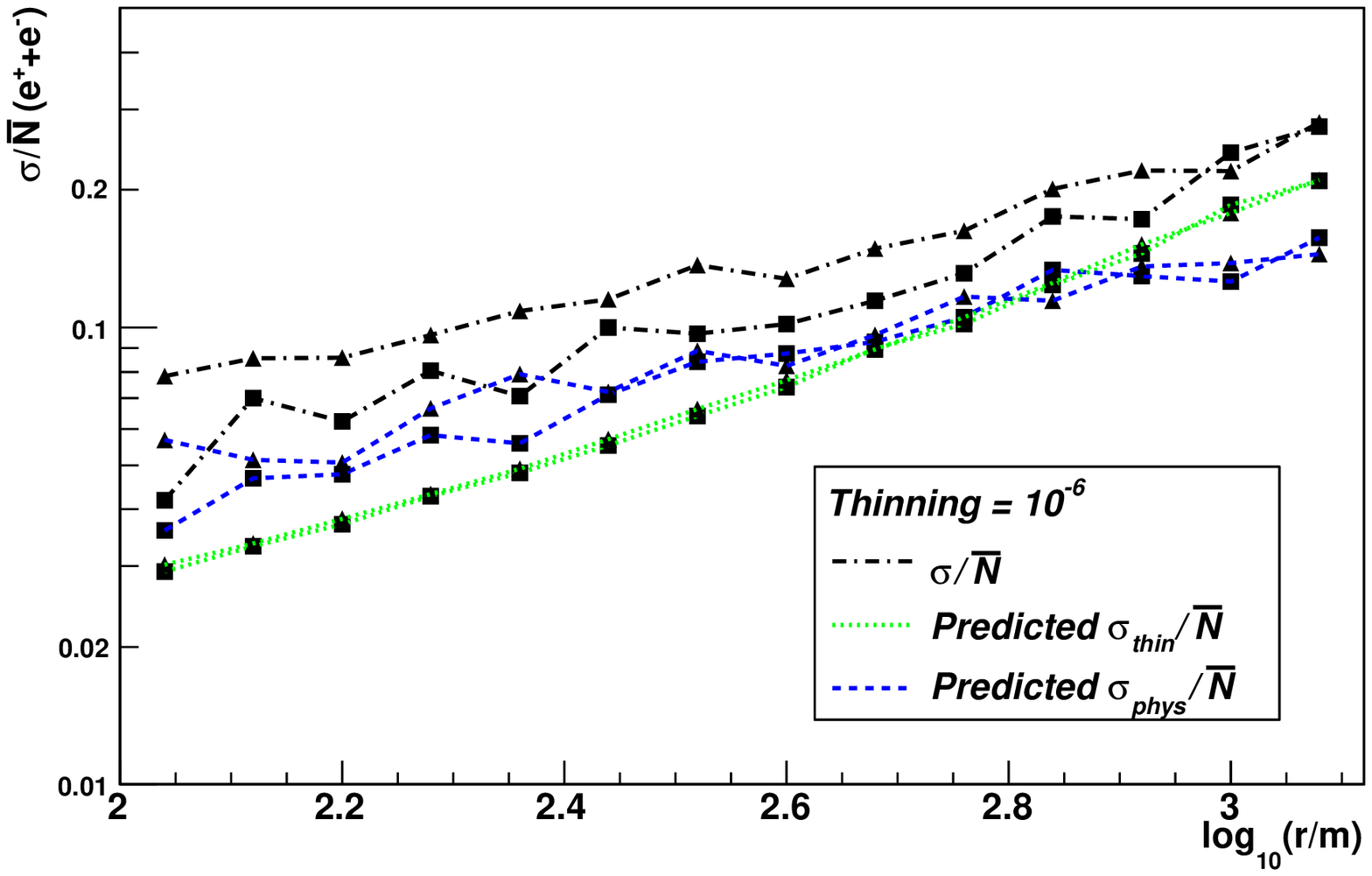}
&\includegraphics[angle=0, width=7.0cm]{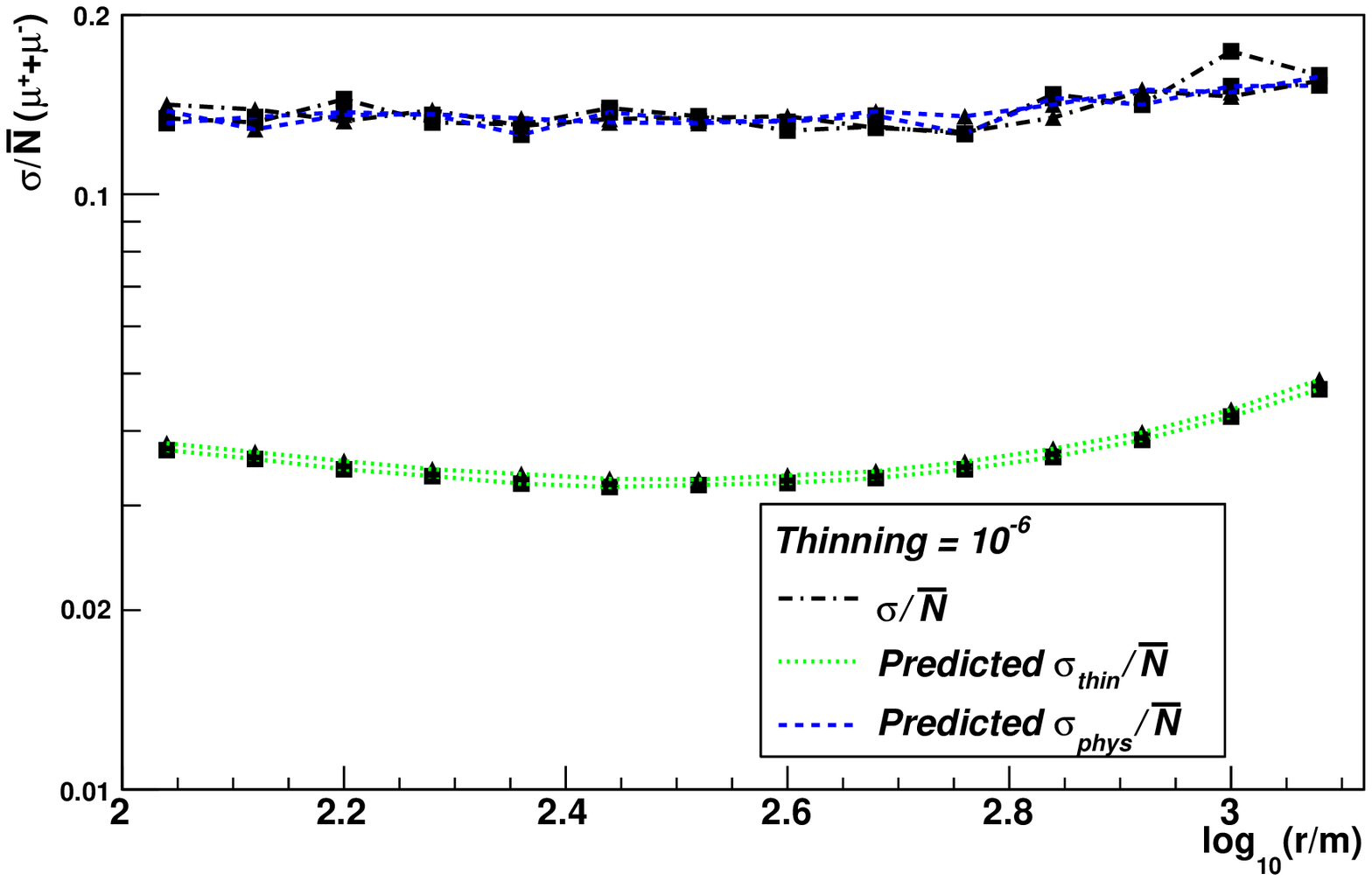}
\\
\end{tabular}\par}
\caption{Relative fluctuations $\sigma/\bar N$ of the distribution of 
number of electron (left panel) and
  muons (right panel) at ground, as a
  function of the logarithm of the distance to the core, for $10^{19}$ eV proton
  showers with $\theta=0^\circ$, and relative thinning level $R_{\rm thin}=10^{-6}$. 
  We show (squares) the result of fixing the first
  interaction depth at 44.9 g/cm$^{2}$ (mean interaction depth of $10^{19}$ eV
  proton-air collisions predicted by the QGSJET01 model), and also the case in which the
  depth of first interaction fluctuates (triangles). The $\sigma/\bar N$ obtained in the simulations is compared to that predicted by 
Eq.~(\ref{eq:sigma}). The two terms in Eq.~(\ref{eq:sigma}) corresponding to fluctuations
induced by thinning $\sigma_{\rm thin}$ and physical fluctutations $\sigma_{\rm phys}$ are 
also shown in all cases, see insets.}
\label{fig:fig13}
\end{figure}
\begin{figure}[ht]
{\centering \begin{tabular}{cc}
\includegraphics[angle=0, width=7.0cm]{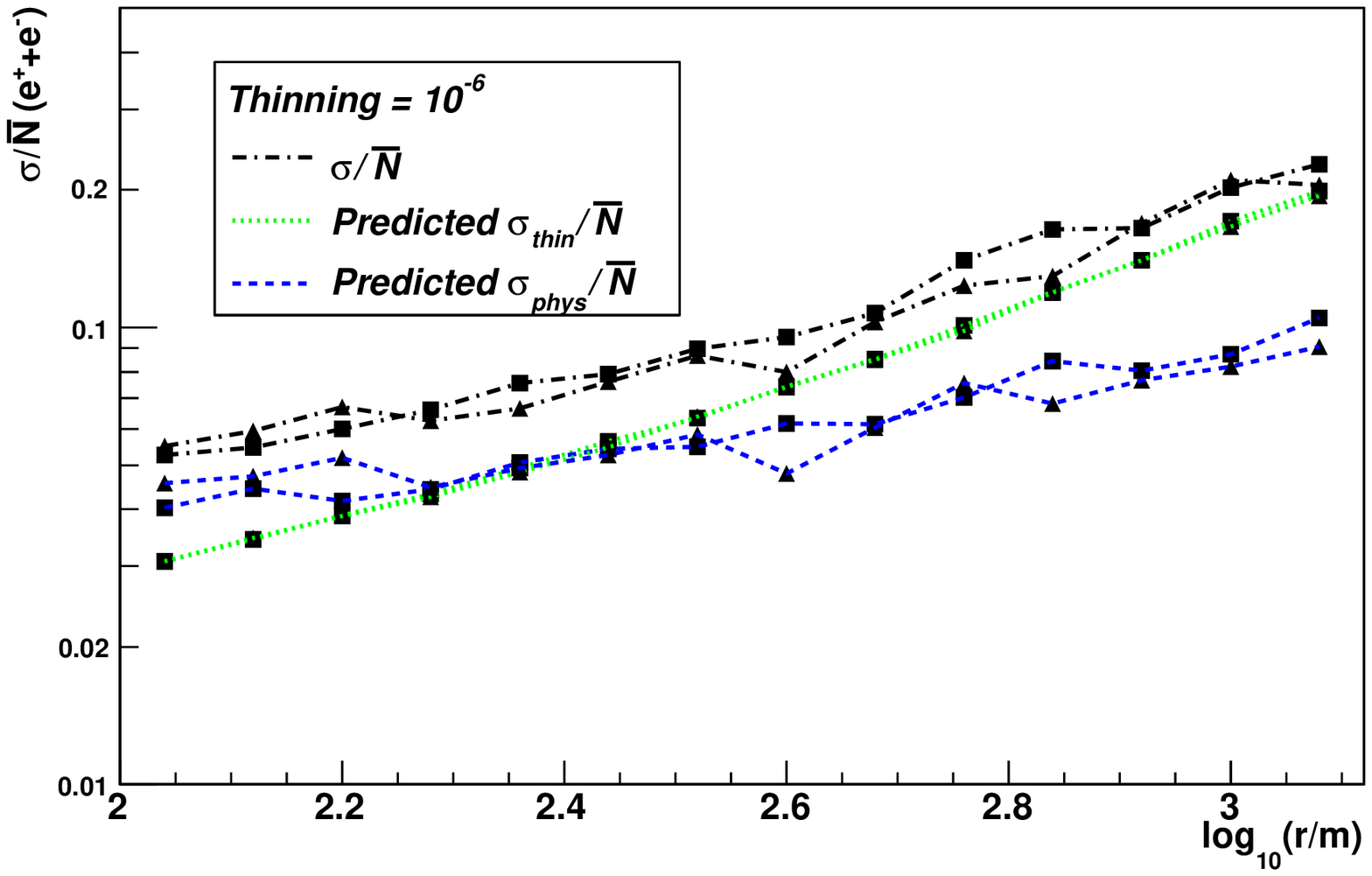}
&\includegraphics[angle=0, width=7.0cm]{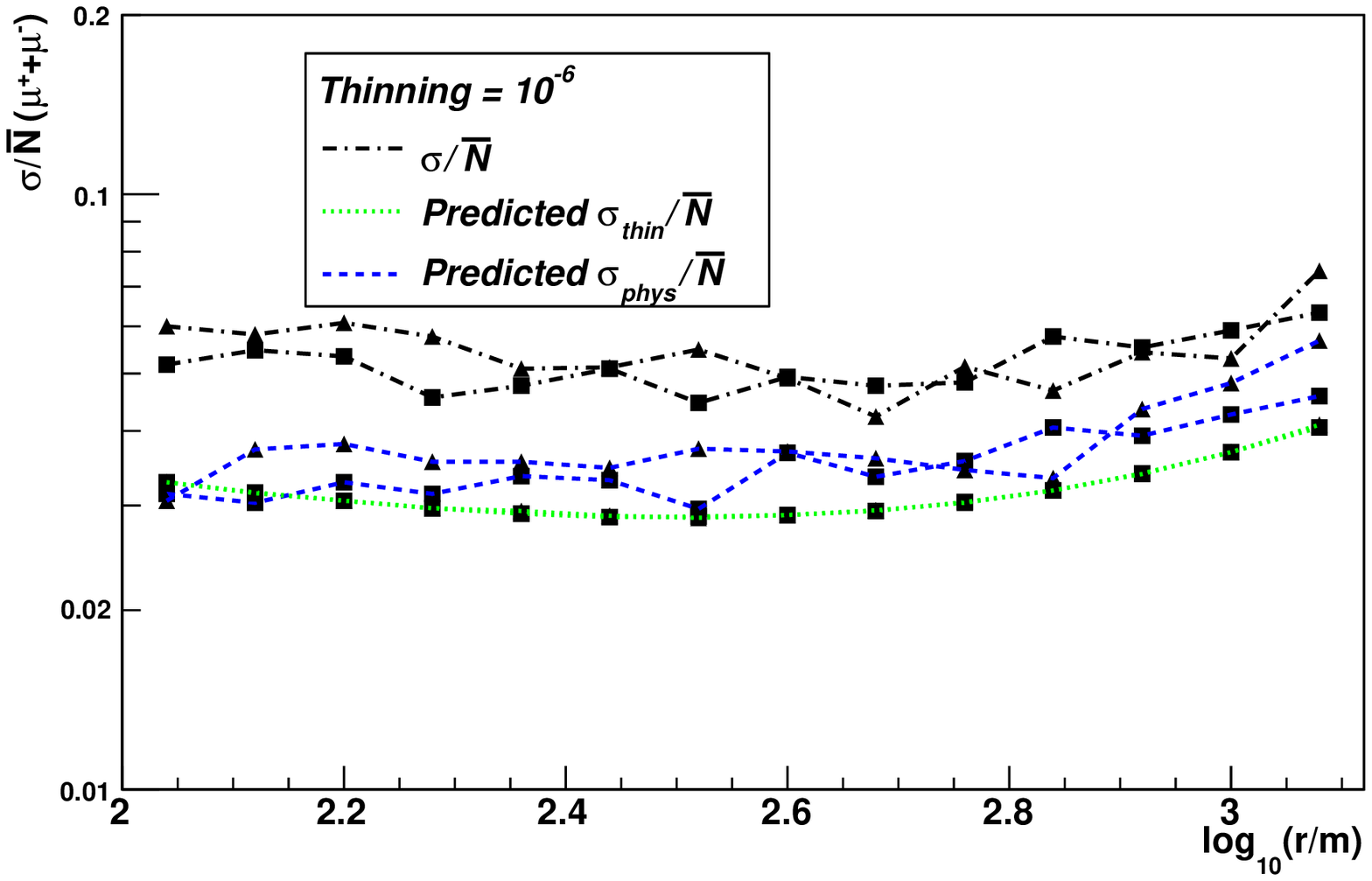}
\\
\end{tabular}\par}
\caption{ Same as Fig.~\ref{fig:fig13} for iron-induced showers. 
In the simulations with fixed first interaction depth, that depth was 
chosen at 10.7 g/cm$^{2}$ (corresponding to the mean
interaction depth of $10^{19}$ eV iron-air collisions predicted by the QGSJET01 model).
Same symbols and line types as in Fig.~\ref{fig:fig13}.
}
\label{fig:fig14}
\end{figure}

\section{Dependence of shower to shower fluctuations on composition and 
depth of first interaction}
\label{S:Composition}

In this Section we study the influence of the fluctuations in the depth
of first interaction on the overall shower to shower fluctuations of the number of 
particles. In Figs.~\ref{fig:fig13} and \ref{fig:fig14} we plot the relative
fluctuations $\sigma/\bar N$ in $10^{19}$ eV proton and iron-induced showers respectively.  
In all panels we show the results of our regular simulations, together with 
the results of a special set of simulations performed by fixing the depth of first interaction of the
primary particle (proton or iron) at the value of its mean interaction depth 
predicted by the QGSJET model (namely 44.9 g/cm$^{2}$ for proton
at $10^{19}$ eV and 10.7 g/cm$^{2}$ for iron at the same energy). 
In all cases we use Eq.~(\ref{eq:sigma}) to split the fluctuations into 
artificial and physical fluctuations, and we also show them in the figures.  

Firstly, it is interesting to see that the artificial fluctuations in the number of electrons
or muons in iron showers are approximately equal to those in proton showers, while
the physical fluctuations are smaller in iron than in proton-induced showers.  
The latter observation is a well-known effect which is attributed to the fact that 
showers initiated by a nuclei can be considered, in a first approximation, as a
superposition of $A$ (atomic mass) nucleons, each with an energy $E/A$ with 
$E$ the energy of the primary nucleus. 

It is rather remarkable that the relative fluctuations in the number of 
particles $\sigma/\bar N$ in the two different sets of simulations 
(fixing or varying the depth of the first interaction point), are essentially
the same. This conclusion applies to both the number of electrons and 
the number of muons. One could think 
that this is due to the fluctuations induced by thinning which mask the effect of the
fluctuations of the depth of first interaction, however  
this does not seem to be the case, since as can be seen in Figs.~\ref{fig:fig13} and
\ref{fig:fig14}, neither
the first term of Eq.~(\ref{eq:sigma}) (the thinning fluctuations), nor the
second term (the physical fluctuations) change much when varying or fixing 
the depth of first interaction. We conclude that
the relative shower to shower fluctuations on the number of particles at 
ground $\sigma/\bar N$ are rather insensitive to the physical fluctuations of the  
depth of first interaction.
This is related to the fact that the maximum of a shower at $\theta=0^\circ$,
where the fluctuations are minimum \cite{Gaisser}, occurs near the ground. For other zenith
angles a small difference appears in the physical fluctuations of the simulations performed with 
fixed and fluctuated first interaction point.

\section{Conclusions}
\label{S:Conclusions}

In this work we have performed a comprehensive study of shower to shower fluctuations
by means of Monte Carlo simulations of extensive air showers. 
An understanding of the shower to shower fluctuations will help to improve 
the interpretation of cosmic-ray data.

We have shown that the
determination of the true, physical shower to shower fluctuations is hampered by 
the thinning procedure necessary to simulate in a practical manner air showers
at EeV energies and above. However, we also show that the artificial fluctuations
induced by thinning ($\sigma_{\rm thin}$) can be identified and splitted from the
physical fluctuations ($\sigma_{\rm phys}$) with the aid of Eq.~(\ref{eq:sigma}),
which we have shown to account for all, true and artificial fluctuations appearing 
in the simulations. Eq.~(\ref{eq:sigma}) reproduces the expectation that as 
the thinning level decreases $R_{\rm thin}\rightarrow 0$, and showers are less thinned,
then the artificial fluctuations decrease, the physical ones become dominant,
and they do not depend on $R_{\rm thin}$. 

Our simulations also indicate that the physical shower to shower fluctuations of the number
of particles at ground behave proportionally to the number of particles $N$, while the
artificial fluctuations are Poissonian, i.e., behave as $\sqrt N$.

Besides, we have shown that the size of the relative fluctuations due to the
depth at which the first interaction initiating the shower occurs, is smaller 
or of the same order as the fluctuations occuring in the subsequent secondary 
interactions in the shower. 

\section{Appendix}
\label{S:appendix}

In this Appendix, we calculate the probability distribution of the number of particles $N$
in a given bin of $r$ distance to shower core, or of energy.
making some simplifying assumptions. 
We want to calculate the probability
distribution of particles, possibly in a given bin of $r$ or of energy. 
We will assume that the probability for an entry (a non-thinned particle) 
to have a weight $w_i$ is given
by $P_w(w_i)$. In addition the number of entries, $N_e$ in a given shower is a
random variable with probability distribution $P_e(N_e)$. Our main simplifying
assumption is the following: we will assume that $P_w$ and $P_e$ are
independent of each other. This assumption is only approximate, because in a
shower simulated with thinning both the entries and the weight assigned to
each particle are controlled by the branching of the shower and therefore they
must be related. However, as we will see a posteriori, the approximation is
good enough for our purposes here, and it serves to clarify the role of
thinning. 

Under this approximation we can write the probability $P(N)$ of having $N$
particles as,
\begin{equation}
P(N) = \sum_{N_e} P_e(N_e) \left[ \int dw_1 \cdots dw_{N_e} \; P_w(w_1) \cdots
  P_w(w_{N_e}) \; \delta(w_1+ \cdots + w_{N_e}-N) \, \right].
\label{factor}
\end{equation}
where the $\delta-$function expresses the constraint that 
the sum of weights is equal to the total number of particles. 

In what follows we evaluate this expression first by making further
assumptions about the shape of $P_e$ and $P_w$, and afterwards in the general case 
using the characteristic function, related to the probability distribution. 
The definitions of cumulants and of the characteristic function can
be found in any text book on statistics, for instance \cite{stat}.

We start with the integral
\begin{equation}
I(N_e,N) = \int  dw_1 \cdots dw_{N_e} \; P_w(w_1) \cdots
  P_w(w_{N_e}) \; \delta(w_1+ \cdots + w_{N_e}-N),
\end{equation}
and introduce the Fourier representation for the delta function. 
\begin{equation}
I(N_e,N) = \int  dw_1 \cdots dw_{N_e} \; P_w(w_1) \cdots
  P_w(w_{N_e}) \;\; \frac{1}{2 \pi} \int dk \; e^{ik (w_1+\cdots+w_{N_e}-N)}.
\end{equation}
Changing the order of integration gives
\[
I(N_e,N) = \frac{1}{2 \pi} \int dk~e^{-ik \,  N}  
\int  dw_1 \; P_w(w_1) e^{ik \, w_1} \cdots \int  dw_{N_e} \; P_w(w_{N_e}) 
e^{ik \, w_{N_e}} \nonumber
\]
\begin{equation}
= \frac{1}{2 \pi} \int dk e^{-ik \,  N} 
\left[  \int dw P_w(w) e^{ik w} \right]^{N_e}.
\label{in}
\end{equation}
To further continue with the evaluation of $P(N)$ we need to make additional
approximations. We assume that $P_w$ is a Gaussian
distribution with average $\bar w$ and rms $\Omega$
\begin{equation}
P_w(w) = A \; e^{-(w-\bar w)^2/(2 \Omega^2)},
\end{equation}
where $A=1/\sqrt{2 \pi \Omega^2}$ is the probability normalization.
Its Fourier transformation is given by
\begin{equation}
\int d w \; e^{i k w} \; \; P_w(w) = e^{i k \bar w} \; e^{-k^2 \Omega^2/2}.
\end{equation}
Therefore,
\begin{equation}
I(N_e,N) = \frac{1}{2 \pi} \int dk \; e^{-i k (N- N_e \bar w)} \; \;  
e^{-N_e k^2 \Omega^2/2},
\label{fourier_in}
\end{equation}
For large values of $\bar N_e$, the sum in Eq.~(\ref{factor}) 
can be approximated by an integral
\begin{equation}
P(N) = \sum_{N_e} \, P_e(N_e) \; I(N_e,N) \; \approx \int dN_e \; P(N_e) 
\; I(N_e,N).
\label{integral}
\end{equation}
If we assume that $P_e(N_e)$ is also a Gaussian with average $\bar N_e$ and standard deviation 
$s$ and inserting Eq.~(\ref{fourier_in}) in Eq.~(\ref{integral}) gives
\begin{equation}
P(N)= \frac{1}{2 \pi} \int dk \; e^{- i k N} \int dN_e \; 
\frac{1}{\sqrt{2 \pi s^2}} \; 
e^{-(N_e-\bar N_e)^2/(2 s^2)} \; e^{i k N_e \bar w} \; e^{-k^2 N_e \Omega^2/2}.
\end{equation}
The integral over $N_e$ can be done analytically
\begin{equation}
P(N) = \frac{1}{2 \pi} \int dk \; e^{- i k N} \; \; \exp [\,i \bar N_e \bar w k
-\frac{1}{2} (\bar N_e \Omega^2 + s^2 \bar w^2) k^2 + {O}(k^3) \,  ].
\end{equation}
where we neglect in the exponential powers of $k$ larger than 2, after applying 
the saddle point approximation. 
We arrive at the final expression
\begin{equation}
P(N) = \frac{1}{\sqrt{2 \pi \sigma^2}} e^{-(N-\bar N)^2/(2 \sigma^2)},
\end{equation}
where
\begin{align}
\bar N & =  \bar N_e \bar w, \nonumber \\
\sigma^2 & =  \bar N_e \Omega^2 + \bar w^2 s^2. \label{avesigma} \\ \nonumber
\end{align}
Notice that the above expressions have the correct asymptotic behaviour. If the
particles have no weight, $\bar w \rightarrow 1$ and $\Omega \rightarrow 0$,
then the average number of particles is equal to the average number of entries
(non-thinned particles) $\bar N = \bar N_e$ and $\sigma = s$. On the extreme case
of a strongly thinned shower in which all particles
are grouped together in a single entry ($\bar N_e = 1, s= 0$) then $\bar N =
\bar w$ and $\sigma = \Omega$, as expected. 

The above result is general and valid for any probability
distribution for $P_w$ and $P_e$. The only requirement is the ``factorization"
property given in Eq.~(\ref{factor}). From Eq.~(\ref{in}), we introduce the
characteristic function for the probability distribution $P_w$, 
\begin{equation}
\tilde P_w(k) = \int dw \; P_w(w) \; e^{i k w},
\end{equation}
which, in general can be written as
\begin{equation}
\tilde P_w(k) = \exp [\; i k a_1 -\frac{1}{2} a_2 k^2 + \cdots ] = e^{i g(k)},
\end{equation}
where the coefficients of the expansion of $g(k)$ are related to the cumulants
of the distribution of $P_w$. For instance $a_1 = \bar w$, $a_2 =
\Omega^2$, etc. Then Eq.~(\ref{factor}) reads
\begin{equation}
P(N) = \frac{1}{2 \pi} \int dk \; e^{-i kN} \int dN_e \; P_e(N_e) \; 
e^{i N_e g(k)}. 
\end{equation}
We define
\begin{equation}
\tilde P_e(q) = \int dN_e \; P_e(N_e) \; e^{i q N_e} = 
\exp [\; i b_1 q -\frac{1}{2} b_2 q^2 + \cdots],
\end{equation}
where as before $b_1 = \bar N_e$ and $b_2 = s^2$. 
Then we get 
\begin{align}
P(N) = & \frac{1}{2 \pi} \int dk \; e^{-i kN} \; \tilde P_e(g(k)) \nonumber \\
= &  \frac{1}{2 \pi} \int dk \; e^{-i kN} \; \exp [\; i b_1 g(k) 
-\frac{1}{2} b_2  g(k)^2 + \cdots ]. \nonumber \\ \nonumber
\end{align}
Where the function $g(k) = k \bar w + i/2 \; k^2 \Omega^2 + \cdots$. Then after
some algebra
\begin{equation}
P(N) = \frac{1}{2 \pi} \int dk \; e^{-i kN} \; \exp [\; i k \bar w \bar N_e -
\frac{1}{2} k^2 (s^2 \bar w^2+ \bar N_e \Omega^2) + \cdots \; ]
\end{equation}
The coefficients of the expansion around $k=0$ are again the cumulants of the
distribution $P(N)$, therefore we simply read the result given above in 
Eq. \ref{avesigma}. But, as a bonus, we obtain also all the other
cumulants. For instance for the skewness we obtain
\begin{equation}
\gamma_3 = \frac{M_3}{\sigma^3} = \frac{1}{\sigma^3}\times 
(m_3 \bar N_e + 3 \bar w s^2 \Omega^2 + \bar w^3 M_e),
\end{equation}
where $m_3$ is the third central moment ( $m_3 = \langle (w- \bar w)^3)
\rangle \; $) 
of the weight distribution and $M_e$ is the third central moment of the distribution 
of the number of entries. In the same way, one can easily obtain other
cumulants from the above expressions. 

\section{Acknowledgements}
We thank V. Canoa, G. Rodriguez--Fernandez, T. Tarutina, I. Vali\~no and E.
Zas for discussions and comments. P. M. H. was supported by Juan de la Cierva
grant. We thank Centro de Supercomputaci\'on de Galicia (CESGA) for computer
resources.  This work was made possible with support from the Ministerio de
Ciencia e Innovaci\'on, Spain under grant FPA 2007-65114 and Consolider CPAN; 
and of ALFA-EC funds in the framework of the HELEN (High Energy Physics Latin-American-European Network)
project. J. A-M also thanks Xunta 
de Galicia (INCITE09 206 336 PR) for financial support.

\end{document}